\documentclass[
 amsmath,
 aps,
 prx,
 twocolumn,
 superscriptaddress
]{revtex4-2}

\usepackage{xspace}
\usepackage[colorlinks]{hyperref}

\usepackage{gensymb}
\usepackage{float}
\usepackage{graphicx}
\usepackage{epstopdf}
\usepackage{setspace}
\usepackage{bbold}
\usepackage{comment}
\usepackage{color}
\usepackage{soul}
\usepackage[abs]{overpic}
\usepackage{amsmath}
\usepackage{lipsum}
\usepackage[abs]{overpic}
\usepackage{nicefrac}
\usepackage{graphicx}
\usepackage{dcolumn}
\usepackage{bm}
\usepackage{feynmp-auto}

\newcommand{\abs}[1]{\left\vert#1\right\vert}

\begin{document}

\title{Search for Dark Matter Scattering from Optically Levitated Nanoparticles}
\author{Yu-Han Tseng}
\affiliation{Wright Laboratory, Department of Physics, Yale University, New Haven, Connecticut 06520, USA}
\author{T. W. Penny}
\affiliation{Wright Laboratory, Department of Physics, Yale University, New Haven, Connecticut 06520, USA}
\author{Benjamin Siegel}
\affiliation{Wright Laboratory, Department of Physics, Yale University, New Haven, Connecticut 06520, USA}
\author{Jiaxiang Wang}
\affiliation{Wright Laboratory, Department of Physics, Yale University, New Haven, Connecticut 06520, USA}
\author{David C. Moore}
\affiliation{Wright Laboratory, Department of Physics, Yale University, New Haven, Connecticut 06520, USA}
\affiliation{Yale Quantum Institute, Yale University, New Haven, Connecticut 06520, USA}

\date{\today}

\begin{abstract}
The development of levitated optomechanics has enabled precise force sensors that operate in the quantum measurement regime, opening up unique opportunities to search for new physics whose weak interactions may have evaded existing sensors.
We demonstrate the detection of impulsive forces acting on optically levitated nanoparticles, where the dominant noise source is provided by measurement backaction.
Using these sensors, we search for momentum transfers that may originate from scattering of passing particlelike dark matter. 
For dark matter that couples to Standard Model neutrons via a generic long-range interaction, this search constrains a range of models in the mass range $1$--$10^7~\mathrm{GeV/}c^2$, placing upper limits on single neutron coupling strength as low as $\leq 1 \times 10^{-7}$ at the 95\% confidence level.
We also demonstrate the ability of using the inherent directional sensitivity of these sensors to separate possible dark matter signals from backgrounds.
Future extensions of the techniques developed here can enable searches for light dark matter and massive neutrinos that can reach sensitivity several orders of magnitude beyond existing searches.
\end{abstract}

\maketitle

\section{Introduction}
Although there is compelling evidence for dark matter (DM) from astrophysical and cosmological observations, the nature of DM is still unknown~\cite{bertone_history_2018, bertone_particle_2005}.
Existing experiments that search for weakly interacting massive particles (WIMPs) or other similar particlelike DM candidates have placed strong constraints on DM-induced nuclear and electronic recoils, but have not yet led to a definitive detection~\cite{lz_wimps_2024, XENON_wimps_2023, PandaX_wimps_2021, DarkSide_wimps_2022, Schumann_DM_review_2019}.
In recent years, the theoretical development of alternative DM models has greatly expanded the parameter space of viable DM candidates, presenting diverse and challenging opportunities to experimentally detect a variety of DM candidates~\cite{Battaglieri_2017, Billard_2021}.

At the same time, advances in levitated optomechanics have opened possibilities for the development of mechanical sensors that could search for previously undetectable forces, including those that may arise from DM~\cite{carney_mechanical_2020, moore_searching_2021}.
Nano- and microparticles can be optically trapped in ultrahigh vacuum, where the center-of-mass motion of these objects becomes well isolated, 
making them extremely precise force sensors~\cite{ranjit_zeptonewton_2016, monteiro_force_2020, gonzalez-ballestero_levitodynamics_2021, liang_yoctonewton_2023}.
When technical noise sources are mitigated, the force sensitivity possible in high vacuum is ultimately constrained by quantum mechanics, dominated by the light-matter interaction between photons and the trapped object.

Experiments working with optically levitated nanospheres of diameter $\sim 100$~nm have now achieved sufficient isolation and control of their motion to permit ground-state cooling~\cite{delic_cooling_2020, magrini_real-time_2021, tebbenjohanns_quantum_2021, kamba_optical_2022, piotrowski_simultaneous_2022, ranfagni_two-dimensional_2022, dania_high-purity_2024}, and extensions of these techniques have allowed measurement and manipulation of their mechanical degrees of freedom in the quantum regime~\cite{rossi_quantum_2024, kamba_quantum_2025}.
Applications taking advantage of the force sensitivity of these systems include tests of quantum mechanics at macroscopic scales~\cite{PhysRevLett.107.020405, RevModPhys.85.471}, detecting individual nuclear decays~\cite{wang_mechanical_2024} and gas molecule collisions~\cite{barker_collision-resolved_2024}, searches for scattering from particlelike DM ~\cite{monteiro_search_2020, 2025arXiv250311645Q}, millicharged particles~\cite{moore_search_2014, afek_limits_2021}, ultralight DM~\cite{Kilian:2024fsg, Dutta:2025ddv, Li:2023wcb, PhysRevLett.134.251001, Higgins_DM_PhysRevD.109.055024, PhysRevD.110.115029}, new short-range interactions~\cite{PhysRevLett.105.101101, venugopalan_search_2024}, and high-frequency gravitational waves~\cite{PhysRevLett.110.071105, Aggarwal_gravitationalwaves_2022, Carney_grav_wave_PhysRevLett.134.181402}.

In this work, we develop mechanical impulse sensors based on silica nanospheres that are optically trapped in ultrahigh vacuum.
We focus on detecting forces acting on these nanospheres that are exerted over a time period much shorter than their mechanical response time (i.e., ``impulses"), by continuously monitoring the position of the optically levitated sensor. 
As illustrated in Fig. \ref{fig:drawing_brief}, such impulses are the expected signal from particlelike DM candidates interacting with a nanosphere, since the DM velocity ($\sim 10^{-3}$~$c$ for virialized DM, where $c$ is the speed of light) leads to $\lesssim$ps interaction times with the nanoparticle.
While not the focus of this work, the impulse sensors developed here are also relevant to detecting forces from radioactive decays within the particles~\cite{carney_searches_2023, wang_mechanical_2024} or single gas molecule collisions~\cite{barker_collision-resolved_2024}.

To directly characterize the sensor response to such impulses, this work demonstrates \textit{in situ} calibration techniques that allow reconstruction of changes in the momentum of the nanoparticle near the ``standard quantum limit'' for broadband impulsive forces.
These sensors are then applied to search for scattering events that may arise from particle-like DM interacting with the sensor.
With an exposure of approximately $5$~fg-month, this search provides constraints on particlelike DM that couples to neutrons via a generic long-range interaction.
Compared to a previous search based on optically trapped microspheres with size $\approx10$~$\mu$m \cite{monteiro_search_2020}, the $10^{3} \times$ better momentum sensitivity demonstrated in this work significantly extends the reach to parameter spaces for lighter DM masses and lower DM velocities, including DM that may become thermalized through its interactions with the Earth or its atmosphere.

\section{Experimental setup}
\label{sec:experimental_setup}

\begin{figure*}
    \centering
    \includegraphics[width=0.8\textwidth]{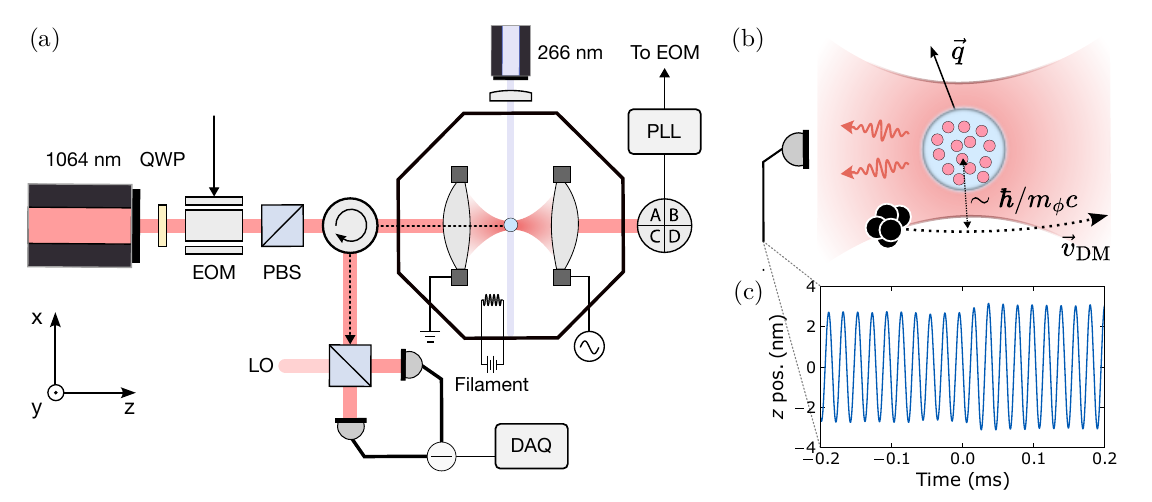}
    \caption{
    Overview of the experimental setup and illustration of the impulse measurement.
    (a) Simplified schematic showing the major features of the experimental setup, as described in the text (see Appendix \ref{sec:detailed_setup} for a full schematic).
    (b) Scattering of a dark matter particle with a levitated nanosphere via a light mediator $\phi$ that couples to neutrons, producing an observable momentum transfer $\vec{q}$. The range of the interaction is approximately $\hbar/m_\phi c$, where $m_\phi$ is the mediator mass. 
    (c) The impulse detection is achieved by continuously monitoring the center-of-mass motion via photons scattered off the nanosphere.
    }
    \label{fig:drawing_brief}
\end{figure*}

Silica nanospheres are trapped in a single-beam optical trap by focusing a 1064~nm infrared laser with $\approx 500$~mW optical power, using an aspheric lens with numerical aperture $\mathrm{NA} =0.77$ in a vacuum chamber (Fig. \ref{fig:drawing_brief}[a]).
The nanospheres have a mass of $m = 4.8 \pm 0.8$~fg, assuming their specified diameter of $166 \pm 9$~nm and density $\rho = 2.0 \times 10^{3}$~kg/m$^3$.
Near the equilibrium position, the center-of-mass motion of the trapped nanosphere is harmonic with frequencies $\Omega_z / 2 \pi = 50 \pm 1$ kHz, $\Omega_y / 2 \pi = 206 \pm 2$ kHz, and $\Omega_x / 2 \pi = 224 \pm 2$ kHz, where the errors represent observed variations between different nanospheres.
The $(x, y, z)$-coordinate system is defined such that the beam propagates in the $+z$ direction and is linearly polarized in the $\pm y$ direction.

After interacting with the nanosphere, the beam that propagates in the $+z$ direction is recollimated by an $\mathrm{NA} =0.5$ lens and directed to balanced photodiodes that detect the nanosphere's center-of-mass motion~\cite{tebbenjohanns_optimal_2019}.
This ``forward detection" signal is used 
to generate feedback that stabilizes the motion of the trapped nanosphere in high vacuum.
Active feedback cooling is implemented by parametrically modulating the power of the trapping laser~\cite{gieseler_subkelvin_2012}.
Photons that backscatter in the $-z$ direction are collimated by the trapping lens and collected by a fiber-based confocal microscope~\cite{vamivakas_phase_2007, magrini_real-time_2021}.
The phase of the backscattered light is read out using homodyne detection, which is primarily sensitive to the $z$-motion of the nanosphere~\cite{tebbenjohanns_optimal_2019, maurer_quantum_2022}.
This ``backward detection" of the $z$-mode provides the main detection channel for the impulse calibration and DM searches presented in Sec.~\ref{sec:calibration} and Sec.~\ref{sec:dm_search}, respectively.

The metallic lens holders in the vacuum chamber are electrically isolated and serve as a pair of electrodes that can produce a controlled electric field at the position of the nanosphere.
The mechanical response of the nanosphere is directly calibrated by applying known voltage pulses to the electrodes, which impart impulses to the nanosphere when it is electrically charged.
In addition, the homodyne $z$-detection is calibrated by applying sinusoidal signals of known frequencies and amplitudes to one of the electrodes.
The electric field at the nanosphere position is modeled using finite-element analysis (FEA), taking into account the detailed geometry of all components surrounding the trap (see Appendix~\ref{sec:app_sub_efield}).
Additional elements in Fig.~\ref{fig:drawing_brief}[a], including a 266~nm pulsed ultraviolet (UV) laser and an electrically heated tungsten filament, are used to control the net electric charge of a trapped nanosphere in ultra-high vacuum, as described in more detail in Sec. \ref{sec:calibration}.

\section{Impulse sensing}
\label{sec:impulse_sensing}
When a nanosphere is optically trapped in vacuum, each of its center-of-mass degrees of freedom can be described as a damped harmonic oscillator interacting with fluctuating forces that arise from random scattering of the trapping photons~\cite{jain_direct_2016, novotny_radiation_2017, abbassi_radiation_2024}, residual collisions with background gas, and any other technical noise sources.
At vacuum pressures $\lesssim 10^{-8}$~mbar, background gas collisions become subdominant, and the interaction of the nanosphere with the trapping and readout photons provides a fundamental source of quantum noise in the measurement of the nanosphere position.
For the laser powers considered in this work, the interaction of the nanosphere with the laser provides a weak continuous measurement of the nanosphere position~\cite{caves_measurement_1980, braginsky_quantum_1992, clerk_introduction_2010}.

For a nanosphere interacting with a trapping laser and coupled to an environmental heat bath via gas collisions, the double-sided, symmetrized power spectral density (PSD) of the homodyne readout signal of the position $\hat{z}$ can be written as~\cite{clerk_introduction_2010, khalili_quantum_2012, whittle_approaching_2021}
\begin{equation}
    \Bar{S}_{zz} =  \abs{\chi}^2 (\Bar{S}^{\textrm{ba}}_{FF} + \Bar{S}^\mathrm{th}_{FF}) + \Bar{S}^{\textrm{imp}}_{zz} +  \Bar{S}^\mathrm{zpf}_{zz},
\label{eq:position_noise}
\end{equation}
where $\chi[\omega] = \left[ m (\Omega^2 - \omega^2 + i \gamma \omega)\right]^{-1}$ is the mechanical susceptibility as a function of angular frequency, $\omega$, for nanoparticle mass $m$, resonant angular frequency $\Omega$, and damping rate $\gamma$.
Here $\Bar{S}^{\textrm{imp}}_{zz}$ is the PSD of the imprecision noise describing uncertainty in the position measurement, $\Bar{S}^{\textrm{ba}}_{FF}$ is the PSD of the backaction force noise, and $\Bar{S}^\mathrm{th}_{FF}$ is the thermal force noise due to residual gas collisions.
$\Bar{S}^\mathrm{zpf}_{zz}$ represents uncertainty due to zero-point fluctuations.

\begin{figure}
    \centering
    \includegraphics[width=0.48\textwidth]{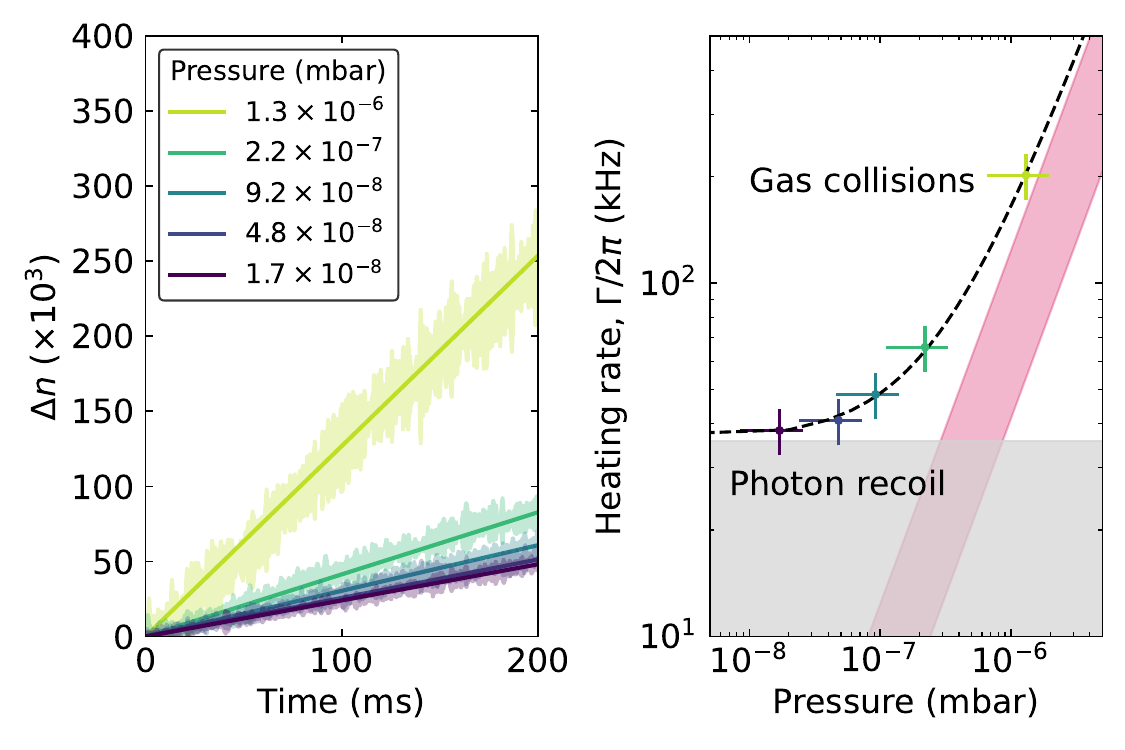}
    \caption{
    Heating rate measurement at various pressures.
    (\textit{Left}) A trapped nanosphere is repeatedly released from feedback at various pressures with its $z$-position continuously monitored.
    The increase in averaged phonon occupancy, $\Delta n \equiv n-n_0$, averaged over 1200 measurements, versus time is shown (plotted in colored bands), with linear fits shown as solid lines.
    (\textit{Right}) Measured heating rates as a function of pressure.
    The horizontal error bars are dominated by systematic uncertainties in the specified accuracy of the pressure gauge.
    The uncertainties of the measured heating rates are dominated by systematic errors in the charge-based position calibration.
    The rates are fit to a model with a constant plus a linear component that represents heating from photon recoil and gas collisions, respectively (dashed line).
    The red band corresponds to the expected contribution of gas collisions (see Appendix~\ref{sec:app_gas}), for residual gas dominated by H$_2$O (upper edge) or H$_2$ (lower edge).
    }
    \label{fig:reheating}
\end{figure}

Forces acting on the nanosphere can be inferred from the position measurement, with corresponding PSD:
\begin{equation}
    \Bar{S}_{FF} = \Bar{S}_{zz} / \abs{\chi}^2.
\label{eq:force_noise}
\end{equation}
For a force $F(t) = \Delta p \cdot \delta(t - t_0)$, where $\delta(t - t_0)$ is the Dirac delta function specifying the impulse time, $t_0$,
the minimum resolvable momentum transfer is~\cite{braginsky_quantum_1992, clerk_quantum-limited_2004, clerk_introduction_2010}:
\begin{equation}
    \label{eq:impulse_integral}
    \Delta p_{\mathrm{min}} =  \left(\int^\infty_{-\infty} \frac{d \omega}{2 \pi} \frac{1}{\Bar{S}_{FF}[\omega]}\right)^{-1/2}.
\end{equation}
For the weak continuous measurements of position considered here, the momentum sensitivity is bounded from below by a ``standard quantum limit'' (SQL) arising from the imprecision and backaction present in the measurement~\cite{braginsky_quantum_1992, clerk_quantum-limited_2004}
\begin{equation}
    \label{eq:impulse_sql}
    \Delta p_{\mathrm{SQL}} = \sqrt{\hbar m \Omega} \approx 24  \ \mathrm{keV/}c  \left(
    \frac{m}{5 \ \mathrm{fg}}\right)^{1/2} \left(\frac{\Omega/2 \pi}{50 \ \mathrm{kHz}}\right)^{1/2}.
\end{equation}
While serving as a convenient benchmark, the impulse SQL in Eq.~\eqref{eq:impulse_sql} is not a fundamental limit and can potentially be surpassed using techniques such as squeezed readout and backaction evasion~\cite{ghosh_backaction-evading_2020, gonzalez-ballestero_suppressing_2023, beckey_quantum_2023, lee_impulse_2025, ghosh_combining_2025}.

\section{Noise analysis}
\label{sec:noise_analysis}
In the absence of technical noise sources, the force noise, $\Bar{S}_{FF}$, consists of thermal noise due to gas collisions and quantum measurement noise, as described in Eqs.~\eqref{eq:position_noise} and \eqref{eq:force_noise}.
To quantitatively estimate the contribution of different noise sources, we measure the heating rate of the $z$-mode at various pressures.
As shown in Fig.~\ref{fig:reheating}, a trapped nanosphere in our system was repeatedly released from feedback cooling for $\approx 500$~ms with its $z$-position continuously monitored.
The averaged phonon occupancy number, $n$, is related to the position measurement by $n = m \Omega^2 \langle z^2 \rangle / (\hbar \Omega)$~\cite{gieseler_dynamic_2014, gieseler_subkelvin_2012, jain_levitated_2017}, where $\langle z^2 \rangle$ is the ensemble averaged variance of position
and $\Omega$ is the resonant angular frequency of the $z$-mode.
The rate of increase in the phonon occupancy measures the strength of the fluctuating force that heats the center-of-mass motion.
When the feedback is turned off, the phonon occupancy of the system initially grows as $n(t) \approx n_0 + \Gamma t$~\cite{gieseler_dynamic_2014, jain_direct_2016, magrini_real-time_2021}, where $n_0$ is the phonon occupancy at time $t=0$ and the total heating rate $\Gamma$ represents the rate at which phonons are introduced into the mechanical mode. 

\begin{figure*}
    \centering
    \includegraphics[width=1\textwidth]{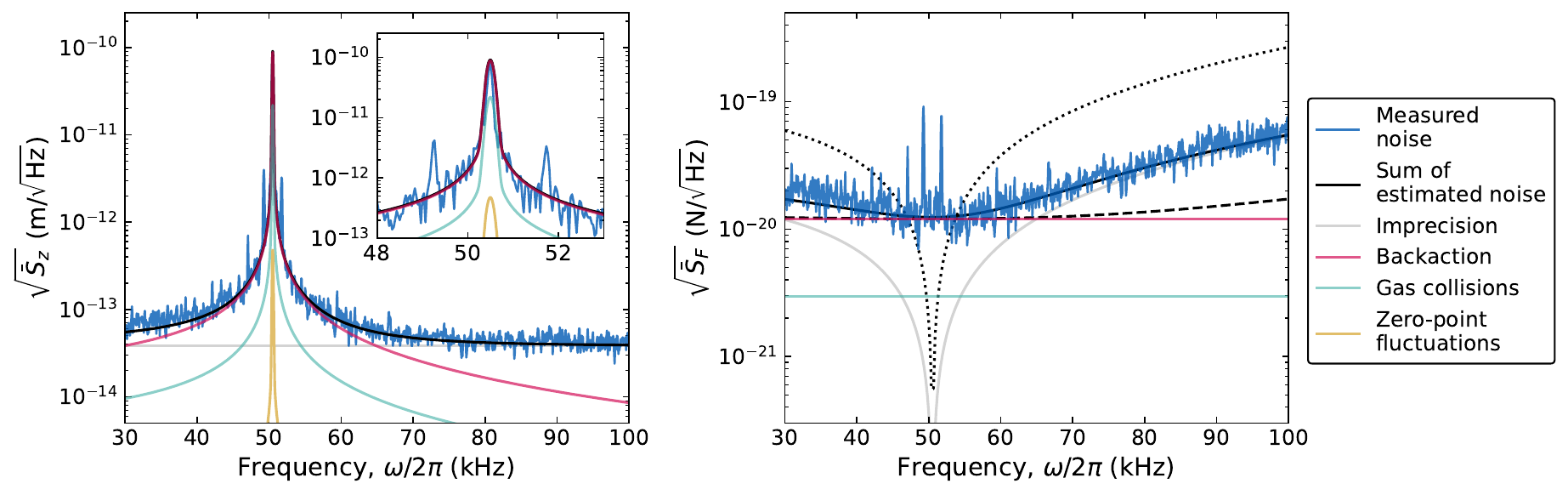}
    \caption{
    Noise spectral densities measured at $(2 \pm 1) \times 10^{-8}$~mbar and the estimated noise budget.
    (\textit{Left}) Single-sided amplitude spectral density of the $z$ displacement noise.
    The contribution of backaction and gas collisions is estimated from the derived heating rate, independent of the position measurement. Narrow features such as vibrational sidebands that are 1.25~kHz away from the resonant frequency are not modeled here.
    (\textit{Right}) Amplitude spectral density of the effective force noise. The contribution of zero-point fluctuations is below the plot range.
    The dashed black line represents the ideal case at the current readout power, in which the imprecision noise is improved by having a measurement efficiency $\eta_m = 1$ and thermal noise eliminated.
    The dotted black line shows the noise spectrum when the impulse SQL (Eq.~\eqref{eq:impulse_sql}) is reached in the idealized situation in which the readout power can be freely tuned without affecting the mechanical response.
    }
    \label{fig:noise_budget}
\end{figure*}

At pressures $\gtrsim 10^{-7}$ mbar, the heating is dominated by random collisions of gas molecules with the nanosphere.
The corresponding thermal heating rate is expected to scale linearly with pressure, consistent with the measurement at high pressures.
When the pressure reaches $\approx 10^{-8}$ mbar, the heating rate is dominated by a pressure-independent force, which we assume arises from random photon scattering~\cite{jain_direct_2016}.
Other technical noise sources are estimated to be subdominant (see Appendix~\ref{sec:app_sub_technical_noises}).
Based on these assumptions, we fit the data to a model $\Gamma = \Gamma_\mathrm{th} + \Gamma_\mathrm{ba}$, where $\Gamma_\mathrm{th}$ is the contribution from gas collisions that is proportional to pressure, while the backaction-induced heating rate, $\Gamma_\mathrm{ba}$, is pressure-independent.
We extract from the fit $\Gamma_\mathrm{ba} = 2 \pi \cdot (35.8 \pm 5.4)$ kHz, in agreement with theoretical expectation (Appendix~\ref{sec:app_backaction_calc}).
The level of backaction force noise is related to the recoil heating rate by $\Bar{S}^{\textrm{ba}}_{FF} = 4 p^2_\mathrm{zpf} \Gamma_\mathrm{ba}$, where $p_\mathrm{zpf} = \sqrt{\hbar m \Omega / 2}$ is the size of the zero-point fluctuations in momentum.
Based on the fit, we estimate the thermal heating rate due to gas collisions to be $\Gamma_\mathrm{th} = 2 \pi \cdot (1.0 \pm 0.2)$ kHz at the minimum operation pressure of $(8 \pm 4)\times 10^{-9}$~mbar.

From the measured displacement noise in the homodyne detection, we estimate the measurement rate $\Gamma_\mathrm{meas} = z^2_{\mathrm{zpf}} / (4 \Bar{S}^{\textrm{imp}}_{zz}) = 2 \pi \cdot (1.9 \pm 0.3$) kHz, where $z_\mathrm{zpf} = \sqrt{\hbar / 2 m \Omega}$ is the size of zero-point fluctuations in position.
The measurement rate represents the rate at which displacement at the scale of the zero-point fluctuations can be resolved.
The corresponding measurement efficiency is $\eta_m = \Gamma_\mathrm{meas} / (\Gamma_\mathrm{ba} + \Gamma_\mathrm{th}) = 0.05 \pm 0.01$.

Fig. \ref{fig:noise_budget} shows the measured noise spectral densities for displacement and force together with the estimated contribution from various sources.
In estimating the effective mechanical susceptibility, a Voigt lineshape plus a constant imprecision noise floor is fitted to the observed displacement noise, $\bar{S}_{zz}$, accounting for broadening due to small drifts in $z$ resonant frequency.
For these measurements, quantum measurement noise sources are found to be the dominant noise in the homodyne readout of the nanoparticle position.
Backaction dominates within $\approx 10$~kHz of the resonant frequency, while imprecision becomes dominant at higher or lower frequencies.
Thermal noise from residual gas is subdominant for the force measurements performed here.
There is evidence of small, unmodeled technical noise at low frequencies.
However, the impact of this noise on our momentum reconstruction is subdominant and can be directly calibrated, and a detailed characterization of its origin is left for future work.

In the weak feedback limit ($\gamma \ll \Omega$) where our system is operated, the broadband integral in Eq.~\eqref{eq:impulse_integral} is dominated by the contribution of backaction around the resonant frequency.
The measured force noise suggests an impulse sensitivity $\Delta p_{\mathrm{meas}} = 65 \pm 9 \ \mathrm{keV/}c$, assuming optimal extraction of the force information from the position measurement. The sensitivity implied by this noise level is within a factor of $\lesssim 3$ of $\Delta p_\mathrm{SQL}$ = 24~keV/$c$.
While stronger feedback cooling would modify the apparent frequency dependence of the contribution from imprecision to the force noise, linear feedback generally offers no advantage in force sensing beyond optimal estimation~\cite{harris_minimum_2013}.

The simple noise model analyzed here suggests a sensitivity $\Delta p = 55~\mathrm{keV/}c$ (Fig.~\ref{fig:noise_budget}, black), which represents the expected sensitivity after residual technical noise is eliminated.
At the current readout power, the sensitivity can in principle reach $\Delta p = 35~\mathrm{keV/}c$ with improved imprecision noise, as indicated by the dashed line in Fig.~\ref{fig:noise_budget}.
To achieve the SQL for impulse sensing in Eq.~\eqref{eq:impulse_sql}, the readout power would need to be further lowered by a factor of $\approx 500$ to reduce the backaction force noise.
However, even in an idealized scenario where the mechanical properties of the oscillator are independent of the readout power, only a small further improvement compared to the current operating condition would be possible due to the weak scaling of $\Delta p$ with readout power~\cite{lee_impulse_2025}.

\section{Impulse calibration}
\label{sec:calibration}
To directly calibrate the nanosphere response, impulses are imparted to a nanosphere with nonzero electric charge using an electric field.
This calibration technique follows similar procedures developed for levitated $\mu$m-size spheres~\cite{monteiro_search_2020, wang_mechanical_2024}, which leverage the precise control of the net electric charge of the trapped object in ultrahigh vacuum.
By directly measuring the response to the forces of interest, this calibration procedure avoids the need to precisely calibrate the displacement sensitivity or mass of the nanosphere.

\begin{figure}
    \centering
    \includegraphics[width=0.5\textwidth]{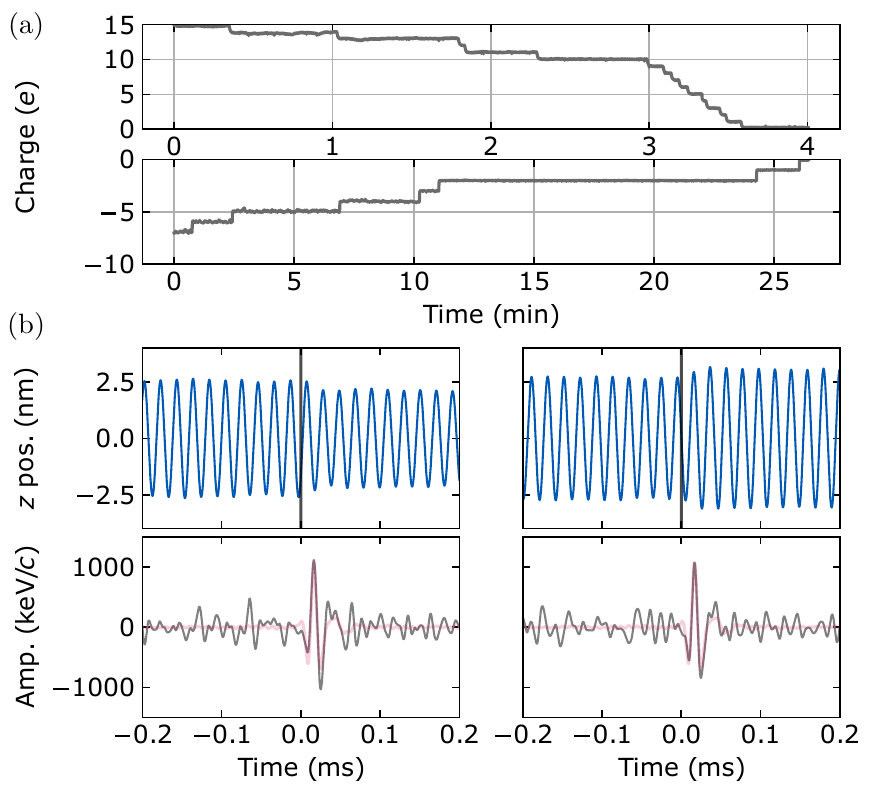}
    \caption{
    Example data demonstrating net charge control in ultrahigh vacuum and impulse calibration.
    (a) Measurement of the net electric charge of a levitated nanosphere as a function of time, with a heated tungsten filament (top) and illumination of a pulsed UV laser (bottom). Discrete steps indicating changes of an elementary charge, $e$, are observed in both cases. The data presented here are taken at a pressure $(2 \pm 1) \times 10^{-8}$~mbar.
    (b) Example position (top panels, blue) and reconstructed impulse amplitude versus time (bottom panels, gray) from electric impulse calibrations. In each calibration, an electric pulse indicated by a solid black line is applied in the $z$-direction at time $t=0$, with an amplitude $\Delta p = 1.1 \pm0.1$~MeV/$c$. The light red lines in the bottom panels represent the reconstructed waveform averaged over 1,200 calibrations.
    }
    \label{fig:charge_impulse}
\end{figure}

Electric charge control of a levitated nanosphere has previously been demonstrated through gas ionization~\cite{frimmer_controlling_2017} and illumination with UV light from a deuterium lamp~\cite{kamba_optical_2022}.
In this work, we extend these techniques to allow controllable tuning of both the charge and its polarity in ultrahigh vacuum. 
Individual electrons can be added to the sphere via thermionic emission of a heated tungsten filament \cite{wang_mechanical_2024}, or removed from the sphere using a pulsed UV laser at a wavelength 266~nm, which is focused onto the trapped nanosphere (see Fig.~\ref{fig:charge_impulse}[a]).
The net electric charge is measured in units of an elementary charge, $e$, by monitoring the amplitude and phase of the motion of the nanosphere when an AC electric drive signal is applied to the $z$-electrodes~\cite{hebestreit_calibration_2018, monteiro_force_2020}.

Once the charge state of a trapped nanosphere is determined, the impulse response of a sphere is calibrated by applying voltage pulses of known amplitudes to the electrodes.
This calibration determines the impulse response to an accuracy for the impulse amplitude $\lesssim 10\%$, where the error is dominated by uncertainty in the knowledge of the electric field at the location of the sphere. 
Other potential sources of systematic error, such as uncertainty in the applied electric waveform and noise due to stray electric field, are estimated to be subdominant.

Compared to the characteristic response time of the oscillator, $\tau \sim 2 \pi / \Omega \approx 20 \ \mu\mathrm{s}$, the force imparted by an electric pulse with a width $\approx 200$~ns is well approximated by an impulse.
A reconstruction algorithm based on matched filtering \cite{golwala_exclusion_2000, wang_mechanical_2024} is used to estimate the amplitude of each pulse (see Appendix~\ref{sec:app_impulse_recon}) \footnote{An alternative approach based on real-time state estimation has been studied in Ref.~\cite{schmerling_optimal_2024}.}.
Examples of reconstructing calibration pulses acting on the sphere are shown in Fig.~\ref{fig:charge_impulse}[b].

\begin{figure*}
    \centering
    \includegraphics[width=0.9\textwidth]{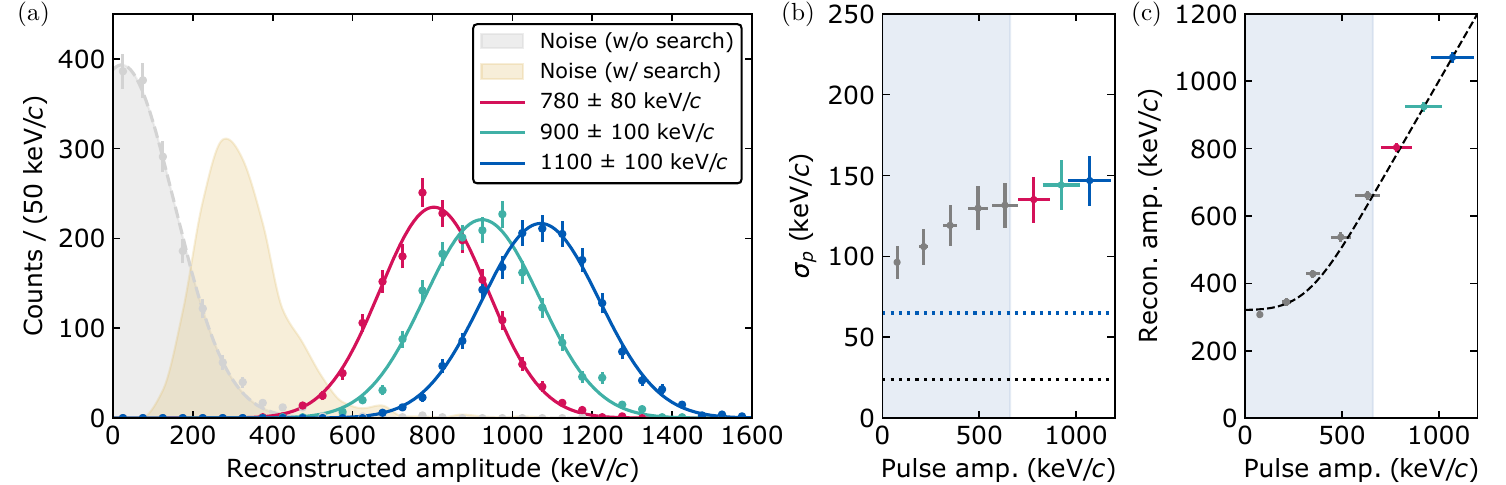}
    \caption{
    (a) Example distributions of the reconstructed amplitudes for applied impulses at various sizes. The solid lines represent Gaussian fits to the measured distributions. The yellow shaded region shows the reconstructed amplitude distribution when the same reconstruction algorithm is applied to noise-only data (i.e., when no calibration impulses are applied). The distribution of the random force amplitudes estimated from the noise-only data is indicated in gray, which represents the force noise level in the absence of search bias, with a Gaussian fit to this distribution shown by the dashed gray line.
    (b) Reconstructed momentum resolution, $\sigma_p$, measured for calibration impulses at different amplitudes. The blue shaded region shows where the impulse amplitude is below $5\sigma_p$, with $\sigma_p = 130 \pm 10~\mathrm{keV/}c$ measured from the noise distribution in the left figure.
    The blue and black dotted lines indicate the impulse sensitivity estimated from the measured force noise in Sec.~\ref{sec:noise_analysis} ($\Delta p_{\mathrm{meas}} = 65 \pm 9 \ \mathrm{keV/}c$) and the impulse standard quantum limit (Eq.~\eqref{eq:impulse_sql}), respectively.
    (c) Average reconstructed amplitude versus true impulse amplitude. The dashed line is a fit to a model showing the linearity of the response above the reconstruction threshold, but accounting for the expected nonlinearity from the search bias for amplitudes $\Delta p \lesssim 500~\mathrm{keV/}c$.
    }
    \label{fig:calibration_sphere20250103}
\end{figure*}

We perform impulse calibrations at various amplitudes for each nanosphere by varying either its net charge or the amplitude of the applied voltage pulses.
The amplitude reconstruction is found to be linear for impulses in the range $500 \ \mathrm{keV/}c \lesssim \Delta p \lesssim 22 \ \mathrm{MeV/}c$, where the upper limit represents the largest impulse amplitude investigated.
For smaller impulses with $\Delta p \lesssim 500 \ \mathrm{keV/}c$, nonlinearity in the reconstruction arises from bias due to searching for the maximum reconstructed signal in the presence of noise~\cite{moore_search_2012} (Fig.~\ref{fig:calibration_sphere20250103}[c]).

For the nanospheres studied in this work, momentum resolutions $\sigma_p = 110$--$150~\mathrm{keV/}c$ were measured for impulses with amplitudes close to the detection threshold ($800~\mathrm{keV/}c\lesssim \Delta p \lesssim 1.1~\mathrm{MeV/}c$).
The resolution is found to grow approximately quadratically with the impulse amplitude, potentially due to small mismatches between the signal templates and the nanosphere response, with resolutions $\sigma_p = 400$--$500~\mathrm{keV/}c$ measured at $\Delta p = 22 \pm 6~\mathrm{MeV/}c$.

Fig.~\ref{fig:calibration_sphere20250103} shows an example of the distributions of reconstructed amplitudes from one nanosphere near the detection threshold.
The resolution measured in direct calibration is consistent with the noise level when no calibration pulses are applied, and is approximately a factor of two higher than the impulse sensitivity estimated from the measured force noise in Sec.~\ref{sec:impulse_sensing}.
Although further work is required to determine the broadening mechanism, improved reconstruction methods that more fully account for time-dependent drifts or imperfections in the signal or noise model could allow further improvement to reach the estimated noise-limited resolution.

\section{Dark matter search}
\label{sec:dm_search}
Using the sensors described in the previous sections, data were acquired to search for impulse events that could arise from scattering of particlelike DM.
Specifically, we consider DM models that couple to neutrons via a generic long-range interaction mediated by a mediator $\phi$ (see Fig. \ref{fig:drawing_brief}[b]).
Such models arise in scenarios where DM resides in a complex dark sector that interacts with particles in the Standard Model via a bosonic force carrier (see, e.g.,~\cite{strassler_echoes_2007, knapen_light_2017, zurek_dark_2025}).
For simplicity we consider only neutron couplings here, although these results could be generalized to an arbitrary coupling to matter.

In the limit where the finite size of a levitated nanosphere can be neglected, the interaction between a pointlike DM particle and the nanosphere can be described by a Yukawa potential with interaction strength and range parametrized by $\alpha$ and $\lambda$, respectively:
\begin{equation}
\label{eq:yukawa_point_like}
    V(r) = \frac{\alpha}{r} e^{- r / \lambda}, \quad \alpha \equiv \frac{(N_\chi g_\chi)(N_n g_n)}{4 \pi} \equiv \alpha_n N_n.
\end{equation}
Here $\lambda \equiv \hbar / m_\phi c$ with $m_\phi$ being the mediator mass.
In Eq. \eqref{eq:yukawa_point_like}, we allow the possibility that the DM particle itself could be composite (with $N_\chi$ constituents), where $g_\chi$ is the coupling strength between $\phi$ and the individual DM constituents.
$N_n$ is the number of neutrons in the sensor and $g_n$ is the strength of $\phi$-neutron coupling.
For convenience, $\alpha_n$ is defined as the total coupling of a DM particle to a single neutron, which includes the number of constituents if $N_\chi > 1$.

A DM particle passing at a velocity $v$ would transfer momentum to the nanosphere over a time $\Delta t \lesssim b_\mathrm{max} / v$, where $b_\mathrm{max}$ is the maximum impact parameter at which an appreciable momentum is transferred.
For the DM parameter spaces considered in this work, the interaction time $\Delta t \sim b_\mathrm{max} / v\ll \tau$, where $\tau \approx 20~\mu s$ is the characteristic response time of the $z$-motion, leading to an interaction well approximated by an impulse.

We analyze data from two different nanospheres, taken from December 2 to December 17, 2024 (``Nanosphere 1'') and January 3 to January 26, 2025 (``Nanosphere 2''), respectively.
The experiment was carried out in a lab space above ground level at Yale University.
During the data acquisition period, a nanosphere was optically trapped in vacuum ($\lesssim 5 \times 10^{-8}$ mbar) with its center-of-mass motion parametrically cooled to an effective temperature $\approx 10$~mK.
The homodyne readout of the backscattered $z$-signal was continuously digitized to monitor the nanosphere position at a sampling rate of 500~kHz and recorded as 60-second long data files, with 2~seconds of deadtime between each file.

To characterize the mechanical response of the trapped nanospheres, regular impulse calibrations were performed every 2--4 days throughout the data taking period.
For each calibration, the sphere was discharged to a net charge of $8~e$ and 200 impulses at each of eight  different amplitudes from 80~keV/$c$ to 1.1~MeV/$c$ were applied, following the procedure described in Sec. \ref{sec:calibration}.
In between calibrations, the electric charge was monitored and kept at a relative low charge state ($\lvert q \rvert \lesssim 10~e$) to minimize uncontrolled electric coupling (although we did not observe excess noise or background events that are correlated to charge; see Appendix~\ref{sec:app_background}).
In addition to measuring the sensor response, these calibrations are used to determine the efficiency for reconstructing signals of a given amplitude. 

The recorded $z$-signal for the DM search is processed following the same reconstruction procedures as the calibration data to search for candidate impulse events.
A trigger-free analysis is performed, where the maximum reconstructed amplitude in each $50~\mu\mathrm{s}$-long search window is determined, with the window length set by the timing resolution for reconstructing multiple pulses.
The corresponding amplitudes are recorded as candidate impulse events, with each candidate waveform centered at the time of the maximum reconstructed amplitude in the search window.
A $\chi^2$ value is also calculated for each candidate waveform, using the average filtered waveform of the calibration impulses as a template to compute the goodness-of-fit to each candidate impulse.

\begin{figure*}
    \centering
    \includegraphics[width=0.9\textwidth]{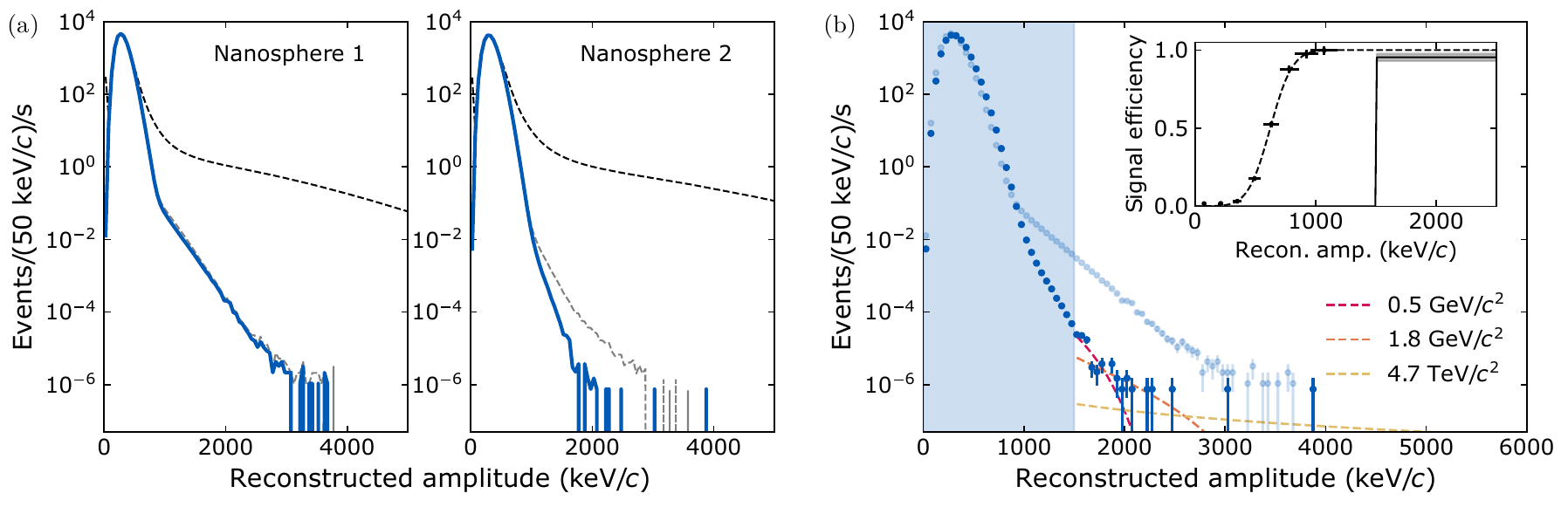}
    \caption{
    (a) Differential event rate measured from the two nanospheres considered in the analysis after all data cuts (blue, solid), with only detection quality and noise level cuts applied (gray, dashed), and with no cuts applied (black, dashed).
    (b) Differential rate measurements after all data cuts, from Nanosphere 1 (light blue) and Nanosphere 2 (dark blue). The dashed lines represent the expected event rate spectra for virialized DM at the 95\% C.L. exclusion limits at each mass. The dark matter sensitivities are dominated by data from Nanosphere 2 because of the significantly lower background rate (see Sec.~\ref{sec:dm_constraints}). Data in the shaded region are below the analysis threshold used when deriving the upper limits. (Inset) Signal efficiency above the analysis threshold after applying all data cuts (black solid line) with the estimated $1\sigma$ uncertainty (gray band).
    Also shown are reconstruction efficiencies near the analysis threshold measured from electric impulse calibrations (black data points) and an error function fit (black, dashed).
    }
    \label{fig:recon_hists}
\end{figure*}

Data selection cuts are applied to remove data containing significant drifts in the detector response or evidence of environmental noise.
First, a detection quality cut rejects any $10~\mathrm{ms}$-long windows in which there is a discontinuous jump in the homodyne detection signal caused by phase-slipping of the local oscillator, removing $4.7\%$ of the total livetime.
A noise level cut is then applied to reject $10~\mathrm{ms}$-windows during which the rms reconstruction noise level is higher than the average noise level by more than $3\sigma$, further removing $8.6\%$ of the livetime.
This cut primarily removes sporadic periods with oscillatory noise consistent with the phase-locked loop in the feedback losing lock.
Finally, data taken from periods in which the background event rates show significant deviation from the steady-state background are excluded from the analysis (see Appendix \ref{sec:app_background}), further removing 7.0\% of the livetime.
The total livetime after all selections is 26.1 days.

Following livetime selection, an event-level cut based on the $\chi^2$ fit to the signal template is applied.
This cut rejects noiselike events whose reconstructed waveforms are inconsistent with the expected shape for an impulse determined from the electric calibrations. 
An amplitude-dependent cut threshold is empirically set to accept an equal fraction of calibration impulses at amplitudes 1--20 MeV/$c$, with the signal efficiency measured to be $95.4 \pm 2.3 \%$ from a dedicated calibration over the full amplitude range considered in the analysis.

The distributions of the reconstructed impulse amplitudes are shown in Fig.~\ref{fig:recon_hists}.
After data selection cuts, the measured event rates below $\approx 1~\mathrm{MeV/}c$ are consistent with the expected Gaussian noise distribution.
Well-reconstructed impulse events above this noise distribution, but with an unidentified origin, were observed in the 1--4~MeV/$c$ range.
No further events between $4~\mathrm{MeV/}c$ and the upper reconstruction threshold of $22~\mathrm{MeV/}c$ were observed after all data cuts.

The observed events were found to have an approximately exponentially falling spectrum that is time-dependent and nanosphere-dependent, but is not significantly correlated to the residual vacuum pressure or the charge state of the nanospheres (see Appendix~\ref{sec:app_background}).
While DM-induced impulses are not expected to show the observed variation in time or between nanospheres, since the origin of these events cannot be definitively identified, we conservatively assume that they could arise either from unmodeled backgrounds or DM scattering when deriving constraints on DM interactions in Sec.~\ref{sec:dm_constraints}.

The impulse measurement demonstrated here is sensitive to both the amplitude and direction of the momentum transfer.
This directionality can be a powerful tool for separating possible DM-induced signals from backgrounds, and confirming the Galactic origin of a signal~
\cite{ahlen_case_2009, mayet_review_2016}.
In particular, if DM particles are virialized within the Galactic halo, the angular distribution of DM-induced recoils is expected to show a periodic modulation due to Earth's rotation with respect to the halo~\cite{spergel_motion_1988, mayet_review_2016}.

While a three-dimensional recoil measurement would be ideal for probing the full structure of the DM velocity distribution~\cite{gondolo_recoil_2002, mayet_review_2016, heikinheimo_probing_2024}, the measurement presented here already provides considerable directional sensitivity.
As a proof-of-principle, we compare the measured event rates to the expected daily modulation signal using data taken from January 3 to January 26, 2025 with Nanosphere 2.
The total livetime for this modulation analysis, after all data selection cuts, is 15.85 solar days (15.89 sidereal days), spanning 22 consecutive sidereal days.
Observed candidate events with amplitudes larger than 1.5~MeV/$c$ are binned in local sidereal time to estimate the event rates after correcting for livetime in each bin.

\begin{figure}[t]
    \centering
    \includegraphics[width=0.5\textwidth]{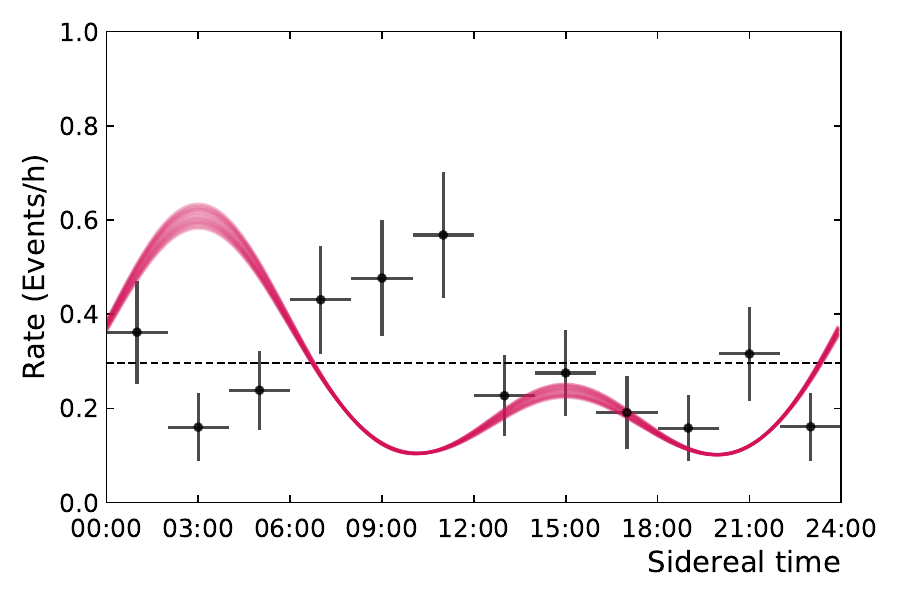}
    \caption{
    Average event rates measured with Nanosphere 2 for impulses with amplitudes larger than $1.5~\mathrm{MeV/}c$, as a function of local sidereal time. The error bars represent the Poisson counting errors.
    Also shown are the average rate (black, dashed) and an example modulation signal from virialized dark matter with $M_\chi=0.5~\mathrm{GeV/}c^2$, $\alpha_n=1.2\times10^{-6}$, and $m_\phi=1~\mathrm{eV/}c^2$ (red band). The dark matter signal is shown as a band because of variation in rate during the data taking period due to Earth's motion with respect to the Galaxy.
    }
    \label{fig:daily_modulation}
\end{figure}

The measured event rate versus sidereal time is shown in Fig.~\ref{fig:daily_modulation} and compared to the modulation signal expected for virialized DM.
In deriving the DM signal, the DM velocity, $\vec{v}$, is assumed to follow the ``Standard Halo Model'' (SHM) where the distribution, $f(\vec{v})$, is a truncated Maxwell-Boltzmann distribution in the Galactic frame, which is then converted to the lab frame~\cite{bozorgnia_daily_2011, mayet_review_2016}.
The resulting DM velocity distribution, together with the numerically calculated DM-nanosphere scattering cross section (see Appendix~\ref{sec:app_dm_rate}), are used to derive the expected event rate for recoils projected onto the $z$-axis.
Other model-dependent effects, including gravitational focusing of DM~\cite{sikivie_solar_2002, alenazi_phase-space_2006} and modification of DM velocities by scattering with matter on Earth~\cite{bertou_earth-scattering_2025} (see also Appendix~\ref{sec:app_overburden}), are not included in this calculation.

Although the observed distribution of events is not consistent with the expected DM-induced rate under the assumptions above, since the observed impulse events may arise from unmodeled time-dependent backgrounds, we do not derive exclusion limits from this modulation analysis and instead consider a more conservative time-independent analysis in Sec.~\ref{sec:dm_constraints}.
However, in future work the unique signature expected for virialized DM in both the modulation amplitude and phase could be used to effectively reject these background events.

\section{Constraints on DM models}
\label{sec:dm_constraints}
For DM particles with mass $M_\chi$ that couple to neutrons via the Yukawa potential in Eq.~\eqref{eq:yukawa_point_like}, the differential rate of DM-induced impulses with respect to momentum transfer $q$ is given by:
\begin{equation}
    \frac{dR}{dq} \left(\alpha_n; M_\chi, m_\phi\right) = n_\chi \int dv\  vf(v) \frac{d \sigma}{dq},
    \label{eq:dm_event_rate}
\end{equation}
where $n_\chi = f_\chi \rho_{\chi} / M_\chi$ is the local DM number density, $f_\chi$ is the fraction of the local DM density that consists of the candidate under consideration (i.e., allowing for only a subcomponent of DM with $f_\chi < 1$ to interact through such a force), and $\rho_\chi$ is the terrestrial DM mass density.
For DM virialized within the Galactic halo, we assume $\rho_\chi = 0.3~\mathrm{GeV/}c^2 \ \mathrm{cm^{-3}}$ \cite{ParticleDataGroup:2024cfk}.
For analyses presented in this section, the DM velocity, $v$, is assumed to have an isotropic distribution, $f(v)$, for which we consider the limiting cases of either virialized DM following the SHM or DM with a thermalized velocity distribution.

To determine the differential cross section $d\sigma / dq$, we numerically calculate the cross section for a pointlike DM particle scattering from a uniform density sphere with a long-range coupling in the form of Eq.~\eqref{eq:yukawa_point_like} (see Appendix \ref{sec:app_dm_rate}).
In the parameter spaces considered here, the de Broglie wavelengths of the DM particles are much smaller than the interaction range, $\lambda \equiv \hbar/m_\phi c$, and the interaction is well described by classical scattering~\cite{xu_resonant_2021}.
After projecting the calculated momentum transfers along the $z$-axis~\cite{gondolo_recoil_2002} and accounting for the measured signal efficiency, the rate is convolved with an amplitude-dependent Gaussian smearing using the calibrated resolution.

The optimum interval method~\cite{yellin_finding_2002, yellin_extending_2007, yellin_ways_2011} is used to determine upper limits on $\alpha_n$, the DM-neutron coupling strength, as a function of DM mass.
Because significantly different event rates are observed for Nanosphere 1 and 2, data are combined using the ``serialization'' method~\cite{akerib_low-threshold_2010, yellin_ways_2011}, with the events observed by each nanosphere ordered separately in $q$.
An ``optimum interval'' in $q$ that most strongly constrains the expected signal is determined from the observed events and is used to derive upper limits on the coupling strength~\footnote{The code used here includes code adapted from the \texttt{DarkLim} package available at \url{https://github.com/spice-herald/DarkLim}.}.
A lower analysis threshold of $1.5~\mathrm{MeV/}c$ is used, which is well above the Gaussian noise distribution in the reconstruction.
The maximum amplitude considered in the analysis determining the exclusion limits is $10~\mathrm{MeV/}c$, which is well below the 22~MeV/$c$ upper threshold over which the reconstruction efficiency and resolution were calibrated.
The measured event rate spectra and signal efficiencies used to derive the exclusion limits presented in the following are shown in Fig.~\ref{fig:recon_hists}[b].

\begin{figure}[t!]
    \centering
    \includegraphics[width=0.5\textwidth]{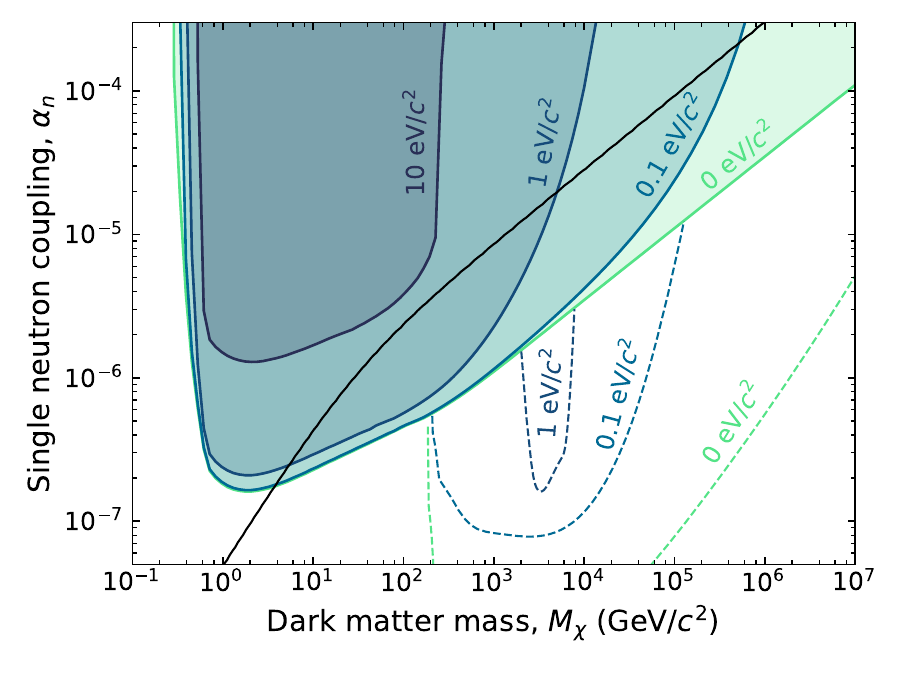}
    \caption{
    95\% C.L. upper limits on the single neutron coupling, $\alpha_n$, for dark matter virialized within the Galactic halo, assuming $f_\chi = 1$ and various mediator masses (colored, solid). Also shown are previous limits from searches using levitated microspheres \cite{monteiro_search_2020} (colored, dashed). The estimated coupling strength above which the dark matter velocity distribution may be significantly affected by the atmospheric overburden is also shown (black, solid).
    }
    \label{fig:combined_alpha_limits_virial}
\end{figure}

There are two main sources of systematic uncertainties associated to the exclusion limits reported here.
First, the masses of the trapped nanospheres are known to within $\approx 20 \%$, and the uncertainty in mass is propagated to the event rate defined by Eq.~\eqref{eq:dm_event_rate} and the derived upper limits.
Systematic errors in the impulse calibration, dominated by the uncertainty in the electric field at the nanosphere position, are propagated as a 10\% uncertainty in the amplitude scale of the observed events.

\subsubsection{Virialized DM}
Fig.~\ref{fig:combined_alpha_limits_virial} shows the 95\% confidence level (C.L.) upper limits on $\alpha_n$ versus $M_\chi$, for various mediator masses.
In deriving these limits, virialized DM with $f_\chi = 1$ whose velocity distribution follows the SHM is assumed.
The upper limits are dominated by the dataset measured with Nanosphere 2 because of its lower background rate (see Fig.~\ref{fig:recon_hists}[b]).
The sub-$\mathrm{MeV/}c$ momentum sensitivity demonstrated in this work probes DM with masses $0.1$--$1~\mathrm{GeV/}c^2$, significantly extending the reach compared to previous searches with levitated microparticles~\cite{monteiro_search_2020}.

Although the SHM is assumed in deriving the upper limits in Fig.~\ref{fig:combined_alpha_limits_virial}, the velocities of incoming DM particles can be significantly modified by scattering of DM with the atmosphere in regions of the parameter space at low mass or large coupling.
The coupling strength above which the velocity distribution of DM particles can show significant deviations from the SHM is indicated by the black solid line in Fig.~\ref{fig:combined_alpha_limits_virial} (see also Appendix~\ref{sec:app_overburden}).
For couplings above this line, the limits are expected to be modified due to the effect of the atmospheric overburden on the DM velocity distribution.
In general, the velocity distribution in such cases can depend on details of the DM model and its accretion history on Earth.
In the following section, the limiting case in which the DM interactions are sufficient to allow it to fully thermalize is presented, demonstrating that the sensors described here can maintain sensitivity to sufficiently heavy DM even at velocities well below the virial velocity.

\subsubsection{Thermalized DM}
In contrast to typical energy-resolving sensors, mechanical sensors can retain significant sensitivity to DM particles as their velocities are reduced well below the virial velocity, since the maximum DM momentum decreases only $\propto v$ (compared to the energy which decreases $\propto v^2$), while at the same time the cross section above threshold can substantially increase. 
In the limit in which the DM-neutron interaction is sufficiently strong, incoming DM particles from the Galactic halo 
continuously lose energy upon entering the atmosphere, and can become gravitationally captured and thermalized~\cite{goodman_detectability_1985, starkman_opening_1990, zaharijas_window_2005, emken_direct_2019}.

\begin{figure}[t]
    \centering
    \includegraphics[width=0.5\textwidth]{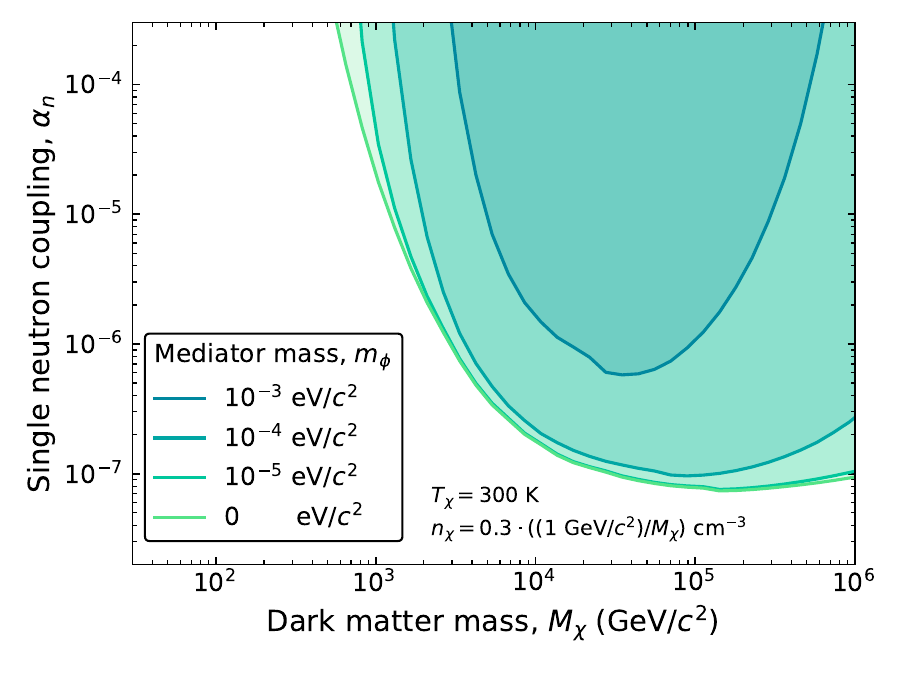}
    \caption{
    95\% C.L. upper limits on the single neutron coupling, $\alpha_n$, for dark matter thermalized to a temperature $T_\chi$ = 300~K and various example mediator masses.}
    \label{fig:thermalized_dm_limits}
\end{figure}

The population of thermalized DM on Earth depends on the assumed DM-matter coupling, the interaction strength, and the complex evolution of astrophysical processes involved in DM capture and accumulation (see, e.g., \cite{neufeld_dark_2018, pospelov_earth-bound_2021, budker_millicharged_2022, leane_floating_2023}).
If such a population of thermalized DM exists, it can evade many existing direct detection searches because of the extremely low energy depositions that would be expected in conventional detectors.
Alternative detector technologies, including trapped ions or electrons~\cite{carney_trapped_2021, budker_millicharged_2022} and superconducting qubits~\cite{das_dark_2024, das_transmon_2024} have been proposed to search for DM of this type.
Here we show that the low-threshold momentum sensing technique demonstrated in this work can also be sensitive to scattering of thermalized DM.

In Fig.~\ref{fig:thermalized_dm_limits}, we show the 95\% C.L. upper limits on $\alpha_n$, assuming DM with a number density at the Earth's surface normalized to the average Galactic halo DM density, $n_\chi = 0.3~(\frac{1~\mathrm{GeV/}c^2}{M_\chi})~\mathrm{cm}^{-3}$.
The velocity distribution of DM is assumed to follow a thermal distribution at a temperature $T_\chi=300~\mathrm{K}$, truncated at Earth's escape velocity (see Appendix~\ref{sec:app_dm_rate}).
All other elements of the analysis remain the same as the virial DM case.

Compared to the case of virialized DM, the behavior of DM-nanosphere scattering changes qualitatively due to the much smaller thermal velocities of DM particles ($v \sim 10^{-6}~c$ for $M_\chi \sim 10^2~\mathrm{GeV/}c^2$).
In the parameter space of interest here, the force imparted by thermalized DM particles remain impulsive.
Our data primarily constrain DM with heavier masses ($M_\chi \gtrsim 10^{3}~\mathrm{GeV/}c^2$) which produce large enough momentum transfers above the analysis threshold.
Due to the increase in cross section at low velocities, in most of the parameter space considered here we find $b_\mathrm{max} \gtrsim \lambda \gg r_\mathrm{sphere}$, where the expected event rates above the analysis threshold are dominated by long-range scattering with impact parameters much larger than the size of the nanosphere.
The DM sensitivities therefore depend strongly on the mediator masses even when $\lambda \gg r_\mathrm{sphere}$.

\subsubsection{Composite DM}
The exclusion limits in Figs.~\ref{fig:combined_alpha_limits_virial}--\ref{fig:thermalized_dm_limits} can also be translated to constraints on specific DM models and compared to existing bounds.
As an example, we consider composite DM (of mass $M_\chi$), in the form of bound states of light constituents (of mass $m_\chi$) that couple to neutrons via a light mediator $\phi$.
These bound states represent a range of DM candidates with long-range self-interactions that can form composite objects consisting of light constituents in a hidden sector~\cite{krnjaic_big_2015, wise_stable_2014, wise_yukawa_2015,  cline_dark_2022}, including specific models such as bound states of asymmetric DM~\cite{gresham_making_2018, coskuner_direct_2019}.

\begin{figure}
    \centering
    \includegraphics[width=0.5\textwidth]{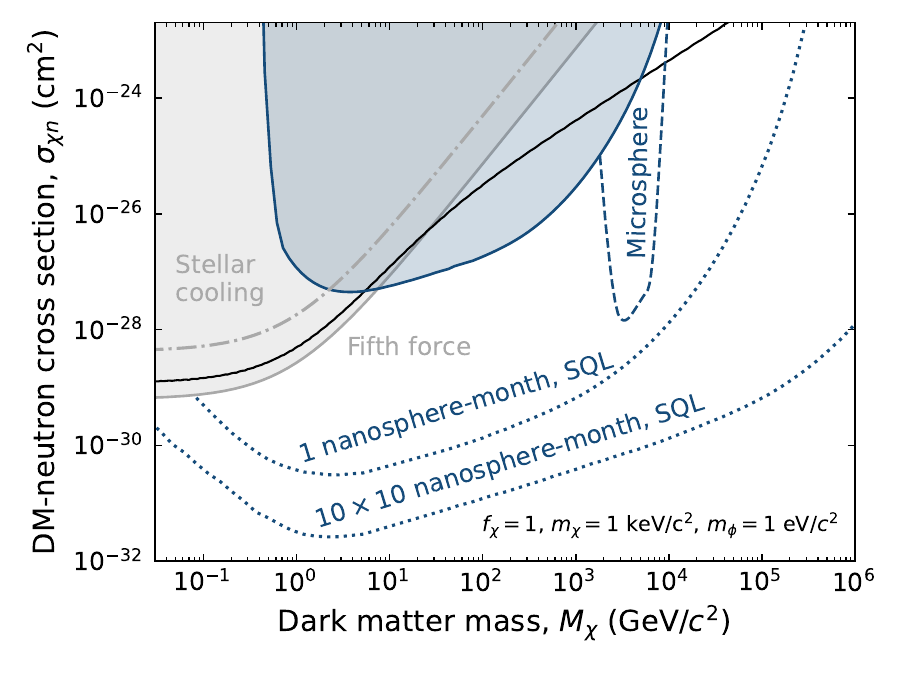}
    \caption{
    95\% C.L. upper limits on the effective dark matter-neutron cross section, $\sigma_{\chi n}$, for virialized, pointlike composite DM for the model parameters described in the text (blue, solid).
    The cross section above which the limits can be significantly affected by atmospheric overburden is shown in black.
    Also shown are limits from microsphere-based searches~\cite{monteiro_search_2020} (dashed) as well as background-free projected sensitivities for a single nanosphere, or an array of such nanospheres, at the standard quantum limit (dotted).
    Model dependent bounds from fifth force experiments~\cite{2015CQGra..32c3001M, lee_new_2020} and stellar cooling processes~\cite{2017JHEP...02..033H} (see also \cite{2023JCAP...07..071B, 2025arXiv250619906F}) are indicated in gray. }
    \label{fig:composite_dm_limits}
\end{figure}

Fig.~\ref{fig:composite_dm_limits} shows the 95\% C.L. upper limits on the effective DM-neutron scattering cross section, assuming composite DM with $f_\chi = 1$, $m_\chi = 1~\mathrm{keV/}c^2$, and $m_\phi = 1~\mathrm{eV/}c^2$.
The effective DM-neutron cross section, $\sigma_{\chi n}$, is defined as $\sigma_{\chi n} \equiv 4 \pi \alpha_n^2 \mu_{\chi_n} / (m_\phi^2 + q_0^2)^2$ in units of $\hbar = c = 1$, where $\mu_{\chi_n}$ is the DM-neutron reduced mass and $q_0 \equiv v_0 \mu_{\chi n}$ is a reference momentum transfer defined at the velocity $v_0 = 10^{-3}~c$~\cite{knapen_light_2017, coskuner_direct_2019, du_atom_2022}.
The strongest constraints on these models come from model-dependent bounds on light mediators from fifth force experiments~\cite{2015CQGra..32c3001M, lee_new_2020} (where we assume a DM-mediator coupling $g_\chi \approx 1$~\cite{knapen_light_2017}) and microsphere-based searches~\cite{monteiro_search_2020} that employ a similar strategy as this work.
Existing direct detection experiments that search for DM-induced nuclear recoils do not significantly constrain these models because of the extremely low expected recoil energy for single nuclei~\cite{coskuner_direct_2019}.

The search presented in this work, with 5~fg of silica and less than a month of exposure, provides sensitivity to these composite DM models that is approximately two orders of magnitude beyond fifth force and stellar cooling constraints for DM masses around 100~GeV/$c^2$.
Future extensions to a large array of levitated sensors~\cite{doi:10.1126/science.abp9941, afek_coherent_2022, siegel_optical_2025} could enable significant improvement beyond this first search employing levitated nanospheres, as shown by the projected reach of a $10 \times 10$ nanosphere array in Fig.~\ref{fig:composite_dm_limits}.

\section{Conclusions}
In this work, we have demonstrated measurement of impulsive forces near the SQL using optically levitated silica nanospheres.
In ultrahigh vacuum, the force noise in the center-of-mass position measurement of these nanospheres is dominated by quantum measurement noise, making them extremely precise force sensors.
Techniques to control the magnitude and polarity of the nanosphere's electric charge in ultra-high vacuum were developed, permitting an \textit{in situ} electric calibration of the impulse response.

We have applied these sensors to search for impulses that may be induced by scattering of particlelike DM, constraining both virialized and thermalized DM with a long-range neutron coupling.
At the current noise and background level, this initial search has already placed constraints for certain DM models, showing the potential of low-threshold optomechanical sensors in DM searches. We have further studied the directional sensitivity of these sensors, indicating that significant additional background rejection is possible.
With a large array of nanospheres and three-dimensional momentum sensitivity, if a positive DM signal were detected, it may be possible to use this directional sensitivity to definitively identify its origin as arising from Galactic DM~\cite{carney_proposal_2020, 2025arXiv250311645Q}.

Future extensions of the impulse measurement technique presented in this work can also find applications in searching for DM interacting coherently with an entire nanoparticle through a heavy mediator~\cite{afek_coherent_2022} or detection of the forces imparted by single neutrinos emitted from within the nanosphere~\cite{carney_searches_2023}.
If the motional state of these levitated sensors can be precisely manipulated at the quantum level, extensions of these techniques may also enable searches for extremely soft scattering of light DM through collisional decoherence~\cite{riedel_direct_2013, riedel_decoherence_2017, du_atom_2022, badurina_coherent_2024}.

\begin{figure*}[!t]
\includegraphics[width=\textwidth]{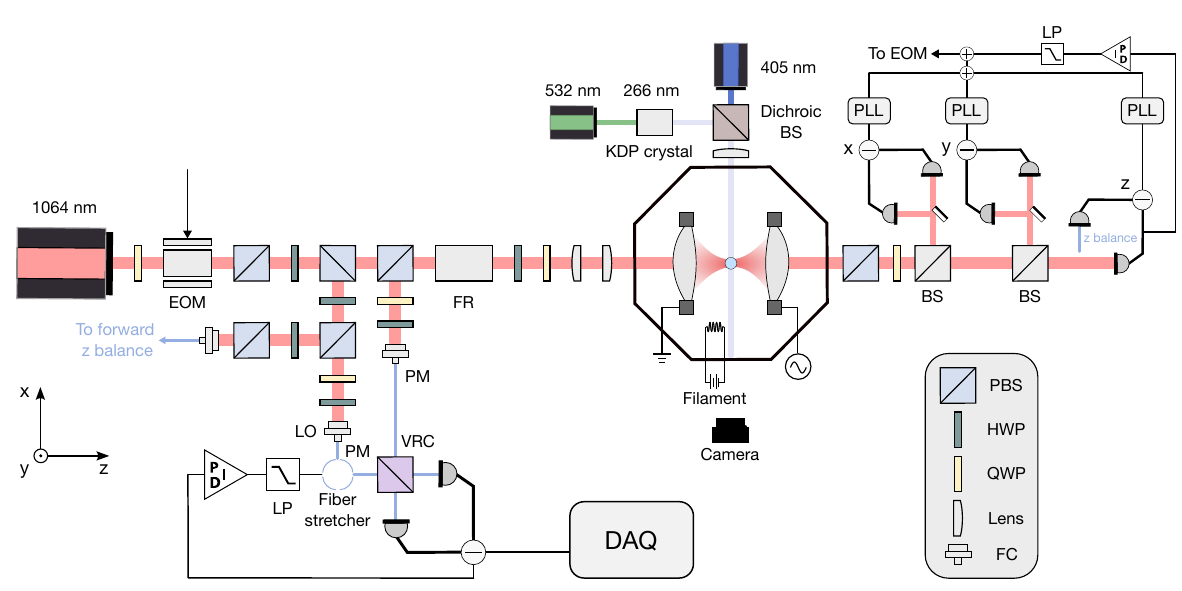}
\caption{Full schematic of the experimental setup, as described in Appendix \ref{sec:detailed_setup}.}
\label{fig:complete_exeprimental_steup}
\end{figure*}

\begin{acknowledgments}
We thank Lorenzo Magrini, Jakob Rieser, Aaron Markowitz, Patrick Maurer, Louisiane Devaud, Xinran Li, Giacomo Marocco, and the QuIPS team at Lawrence Berkeley National Lab for helpful discussions.
We thank Gadi Afek and Geena Benga for their early experimental effort related to this work. This work was supported by NSF Grant Nos. PHY-2512192 and PHY-2109329 and in part by ONR Grant No. N00014-23-1-2600. Y.-H. T. is supported by the Graduate Instrumentation Research Award (GIRA) from the Coordinating Panel for Advanced Detectors (CPAD). 
\end{acknowledgments}


\appendix
\section{Experiment details}
\subsection{Schematic of full experimental setup}
\label{sec:detailed_setup}

Further details of the experimental setup are shown in Fig.~\ref{fig:complete_exeprimental_steup}.
The experiment uses a laser at wavelength $\lambda = 1064$~nm whose intensity is modulated by an electro-optic modulator (EOM) and a polarized beam splitter (PBS).
The beam is rotated to be circularly polarized using a quarter wave plate (QWP) before passing through the EOM.
The beam is then split into two parts using a half wave plate (HWP) and a PBS.
The first part is used to form an optical tweezer through an $\mathrm{NA}=0.77$ aspheric lens in the vacuum chamber, while the other part is used to form the reference beam for forward $z$-detection and the LO for the homodyne backward $z$-detection.
The polarization of the beam forming the optical tweezer is controlled by a HWP and a QWP in front of the vacuum chamber.

The backscattered light from the trapped nanosphere is selected by a Faraday rotator (FR) and a PBS and collected by a fiber collimator (FC) into a polarization-maintaining (PM) single-mode fiber.
The FC and the trapping aspheric lens form a confocal microscope, and the focal length of the FC is selected to optimize the mode-matching between the collected dipole radiation and the optical mode of the PM fiber \cite{magrini_real-time_2021, vamivakas_phase_2007}.
The collected backscattered light is mixed with the LO in a fiber-based variable ratio coupler (VRC) to interferometrically read out the phase of the backscattered light.
The phase of the LO beam is controlled through a fiber stretcher to counter slow drifts due to environmental noise.

The forward scattered light is collected by an $\mathrm{NA}=0.5$ aspheric lens to detect the $x$, $y$, and $z$ motion.
The relatively low $\mathrm{NA}$ of the collection lens is optimized to collect $z$-information that scatters forward in the $+z$ direction~\cite{tebbenjohanns_optimal_2019}.
A PBS and a QWP is used after the vacuum chamber to prevent reflections from coupling into the optical trap.
For the $x$- and $y$-modes, the beam is split by D-shaped mirrors along the two axes and sent to balanced photodiodes to detect the shift of intensity pattern due to the nanosphere's motion in the $x$-  and $y$-directions.
For the $z$-mode, the beam is balanced against a reference beam picked off before the vacuum chamber to measure the intensity modulation due to the $z$-motion.
The $x$, $y$, and $z$ forward detection signals are sent to phase-locked loops (PLLs) to generate feedback that stabilize the nanosphere's motion by parametrically modulating the trapping potential through the EOM, controlled through a \texttt{Red Pitaya}-based field programmable gate array program~\footnote{The source code is available at \url{https://github.com/tefelixhu/redpitayapll} and \url{https://github.com/vhock/Phase-locked-Loop}}.
The measured total beam intensity is also used to generate feedback that corrects for slow drifts in laser intensity and pointing.

As described in Sec.~\ref{sec:calibration}, a pulsed 266~nm UV laser and a tungsten filament are used to control the net charge of the trapped nanosphere.
The 266~nm UV beam is generated by frequency doubling a $532$~nm pulsed laser using a potassium dihydrogen phosphate crystal.
The UV beam is overlaid with a 405~nm laser through a dichroic beam splitter (Dichroic BS) for alignment purposes and focused onto the trapped nanosphere.

\subsection{Electric field simulation}
The impulse calibration described in Sec.~\ref{sec:calibration} relies on knowledge of the electric force that acts on the trapped nanosphere when a voltage signal is applied to the electrodes.
Because the net electric charge is known exactly, this calibration method requires only accurate knowledge of the electric field at the nanosphere position as a function of applied voltage~\cite{monteiro_search_2020, wang_mechanical_2024}.

\begin{figure}
\includegraphics[width=0.375\textwidth]{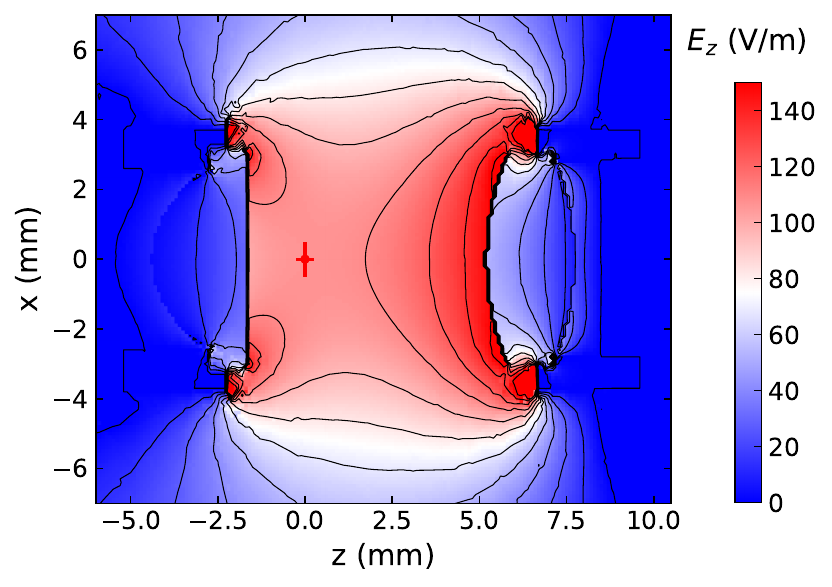}
\caption{Cross-sectional view in the $x-z$ plane of the 3D electric field finite-element simulation.
In the simulation, the $+z$ electrode is kept at $+1$~V while the $-z$ electrode and all other surrounding surfaces are grounded.
The expected sphere position is indicated by the red marker with error bars.}
\label{fig:efield_sim}
\end{figure}

The electric field near the nanosphere position is calculated using FEA with the COMSOL Multiphysics software.
The exact three-dimensional computer-aided design (CAD) model is used in the FEA simulation, which takes into account effects from all the surrounding surfaces and the detailed geometry of the aspheric lenses and lens holders.
Fig. \ref{fig:efield_sim} shows the calculated electric field in the $z$-direction when $+1$~V is applied to one of the electrodes with all other metal surfaces grounded.

To estimate systematic errors from uncertainties in the electrode geometry and nanosphere position, we performed FEA simulations with modified CAD models that represent possible deviations from the original model due to misalignment.
Other sources of systematic error, such as stray electric field from charged dielectrics, are estimated to be subdominant.
Throughout this paper, the electric field in the $z$-direction at the nanosphere position is estimated to be $E_z = 106 \pm 11~ \mathrm{V/m}$ when $+1$~V is applied to the $+z$ electrode, where the error is conservatively estimated based on FEA simulations for extreme cases of deviation.
Based on this electric field simulation, the amplitudes of the calibration impulses are determined by integrating the measured voltage waveforms from the pulse generator that are applied to the electrodes.

\label{sec:app_sub_efield}

\section{Additional details on noise analysis}
\label{sec:app_noise_estimate}
In the noise analysis presented in Sec.~\ref{sec:noise_analysis}, we estimate the contribution of backaction and gas collisions to the observed force noise from direct measurement of heating rate.
Here we provide theoretical estimates and discuss contribution from various noise sources.

\subsection{Backaction (photon recoil)}
\label{sec:app_backaction_calc}
We assume a spherical nanoparticle with a density $\rho = 2,000$~kg/m$^3$ and a diameter 166~nm.
The nanoparticle is modeled as a subwavelength dielectric particle, for which the real part of polarizability is given by $\alpha' = 3 V \epsilon_0 (\epsilon - 1) / (\epsilon + 2)$ with $V$ being the volume of the nanoparticle, $\epsilon_0$ the vacuum permittivity, and $\epsilon \approx (1.4)^2$ the relative permittivity (here we assume a refractive index $n\approx1.4$ specified by the supplier).
The total power scattered by the nanoparticle is given by $P_\mathrm{scatt} = I_0 \sigma$,
where $I_0$ is the tweezer intensity and $\sigma = \frac{8 \pi}{3} (\frac{\alpha' k^2}{4 \pi \epsilon_0})^2$ is the total scattering cross section ($k=2 \pi / \lambda$ is the wave number).
Near the focus, the tweezer can be approximated by a Gaussian beam with an intensity $I_0 = \epsilon_0 c E^2_0 / 2 = 2 P_0 / (\pi w_x w_y)$, where $P_0$ is the total optical power, $E_0$ is the electric field amplitude at focus, and $w_{x,y}$ are the effective beam waists in the $x$- and $y$-directions.

The trap stiffness in $z$, which is determined by the gradient force experienced by the nanoparticle, is approximately $k_z = m \Omega^2 \approx \alpha' E^2_0 / 2 z^2_0$~\cite{Novotny_Hecht_2012, gieseler_dynamics_2014, jain_levitated_2017},
where $m$ is the mass of the particle, $\Omega$ is the resonant angular frequency, and $z_0$ is the effective Rayleigh range around the focus of the trapping lens.
We numerically calculate the electric field configuration for a beam tightly focused by an NA=0.77 lens and determine $w_x$, $w_y$, and $z_0$~\cite{Novotny_Hecht_2012}.
For our experimental configuration with $\Omega / 2 \pi = 50~\mathrm{kHz}$, the calculation predicts $P_\mathrm{scatt} = 41.0~\mathrm{\mu W}$.
The expected double-sided PSD for the backaction force noise is given by~\cite{tebbenjohanns_optimal_2019, seberson_distribution_2020, maurer_quantum_2022}
\begin{equation}
    \Bar{S}^{\textrm{ba},~\mathrm{theory}}_{FF} = (A^2 + \frac{2}{5}) \frac{\hbar k P_\mathrm{scatt}}{c} = 9.3 \times10^{-41}~\mathrm{N}^2/\mathrm{Hz},
    \label{eq:backaction_force_noise}
\end{equation}
where $A \approx 0.83$ is a geometrical factor that depends on the trapping NA~\cite{tebbenjohanns_optimal_2019}.
The corresponding heating rate due to backaction is then~\cite{seberson_distribution_2020, maurer_quantum_2022}
\begin{equation}
    \Gamma^\mathrm{theory}_{\mathrm{ba}} = \frac{z^2_\mathrm{zpf}}{\hbar^2} \cdot \Bar{S}^{\textrm{ba},\mathrm{theory}}_{FF} = 2 \pi \cdot 46.2~\mathrm{kHz}.
\end{equation}
This estimated rate is in agreement with $\Gamma_\mathrm{ba}$ experimentally determined in Sec.~\ref{sec:noise_analysis}, taking into account the $\approx 20\%$ systematic uncertainty in the volume of the nanoparticle.

\subsection{Gas collisions}
\label{sec:app_gas}
Here we estimate the force noise and heating rate arising from collisions of residual gas with the nanoparticle.
The viscous force acting on a spherical object due to gas collisions has been extensively studied based on the kinetic theory of gases~\cite{epstein_resistance_1924, Beresnev_Chernyak_Fomyagin_1990, cavalleri_gas_2010, martinetz_gas-induced_2018}.
However, the situation is complicated for an optically levitated nanoparticle because the surface temperature of the particle in vacuum could be higher than the background gas due to optical absorption, which leads to modified gas dynamics~\cite{millen_nanoscale_2014, hebestreit_measuring_2018}.

Following the analysis of Ref.~\cite{millen_nanoscale_2014, hebestreit_measuring_2018, PhysRevLett.132.133602}, we consider a two-population model in which the residual gas at the impinging temperature $T_\mathrm{im}=295~\mathrm{K}$ interacts with the nanoparticle with a surface temperature $T_\mathrm{surf}$ through either specular or diffuse scattering~\cite{epstein_resistance_1924, corson_calculating_2017, barker_collision-resolved_2024}.
The gas molecules that scatter diffusely re-emerge from the surface of the nanoparticle with an emerging temperature $T_\mathrm{em} = T_\mathrm{im} + \alpha_G (T_\mathrm{surf} - T_\mathrm{im})$, where the accommodation factor $\alpha_G$ describes the energy transfer between the gas and the nanoparticle.

In the free molecular flow regime where the mean free path of the gas is much larger than the size of the nanoparticle, the gas-induced damping is approximately~\cite{millen_nanoscale_2014, hebestreit_measuring_2018, PhysRevLett.132.133602}
\begin{equation}
    \gamma_\mathrm{th} = \gamma_\mathrm{im} + \gamma_\mathrm{em} \approx \frac{32}{3} \left(1 + f \sqrt{\frac{T_\mathrm{em}}{T_\mathrm{im}}} \frac{\pi}{8}\right) \frac{r_\mathrm{sphere}^2 P}{m \bar{v}_\mathrm{gas}},
\end{equation}
where the second term in the parentheses indicates the contribution of diffusive scattering, which occurs with a relative frequency $f$.
Here $r_\mathrm{sphere}$ and $m$ are the radius and the mass of the nanoparticle, $P$ is the vacuum pressure, and $\bar{v}_\mathrm{gas} = \sqrt{8 k_\mathrm{B} T / \pi m_\mathrm{gas}}$ is the mean velocity of the impinging gas with $k_\mathrm{B}$ being the Boltzmann constant and $m_\mathrm{gas}$ the mass of the gas particles.

The double-sided PSD for the force noise in the classical limit is~\cite{millen_nanoscale_2014}
\begin{equation}
    \Bar{S}^{\textrm{th},~\mathrm{theory}}_{FF} = 2 m k_\mathrm{B} (T_\mathrm{im} \gamma_\mathrm{im} + T_\mathrm{em} \gamma_\mathrm{em}),
    \label{eq:thermal_force_noise}
\end{equation}
with the corresponding heating rate $\Gamma^\mathrm{theory}_{\mathrm{th}} = z^2_\mathrm{zpf}/{\hbar^2} \cdot \Bar{S}^{\textrm{th},\mathrm{theory}}_{FF}$.
In the calculation, we assume values $\alpha_G = 0.65$, $T_\mathrm{surf} = 1000~\mathrm{K}$~\cite{hebestreit_measuring_2018}, and $f=0.9$~\cite{epstein_resistance_1924, PhysRevLett.132.133602}.
The heating rate calculated from Eq.~\eqref{eq:thermal_force_noise} is shown in Fig.~\ref{fig:reheating}, which is in good agreement with the measured values.
We note that while an elevated surface temperature of the nanoparticle at $\approx 1000~\mathrm{K}$ agrees with the measurements in Sec.~\ref{sec:noise_analysis} and Ref.~\cite{hebestreit_measuring_2018}, this assumption predicts a factor of $\approx 1.8$ higher heating rate compared to the room temperature case and might require further study.

\subsection{Other technical noise}
\label{sec:app_sub_technical_noises}
Because the trap stiffness of the optical potential is proportional to the laser power, fluctuations in laser intensity lead to parametric modulation of the trapping potential that heats the nanoparticle.
The heating rate due to intensity fluctuation is approximately~\cite{PhysRevA.56.R1095, jain_direct_2016, jain_levitated_2017}
\begin{equation}
\label{eq:rin_heating}
    \Gamma_\epsilon \approx \pi \Omega^2 S^\epsilon n,
\end{equation}
where $S^\epsilon$ is the relative intensity noise (RIN) around two times the resonant frequency and $n$ is the phonon occupancy of the mechanical mode.
Using $S^\epsilon \approx -140~\mathrm{dB/Hz}$ from direct measurement, $\Omega/2 \pi = 50~\mathrm{kHz}$, and $n \approx 10000$, Eq.~\eqref{eq:rin_heating} predicts $\Gamma_\epsilon \approx 2 \pi \cdot 4.9~\mathrm{Hz}$, which is negligible compared to photon recoil heating.

In addition, the nanoparticle is electrically charged during the measurement and is therefore susceptible to electric field fluctuations at the trapping position.
Heating induced by electric field noise has been shown to limit the performance of trapped-ion based quantum systems and is a topic of active research.
The electric field noise in a system is typically geometry-dependent, with several scaling laws in frequency and distance observed.
It is therefore difficult to obtain an accurate estimate of the effect from first principles, or from direct extrapolation of existing measurements to the system considered here.
In the following, we discuss potential sources of electric field noise and establish an empirical upper limit on the noise level in our experiment.

The electrodes surrounding the particle were electrically grounded during the heating rate measurement.
However, heating due to fluctuations in the voltage across the electrodes could arise from the Johnson-Nyquist noise induced by thermal motion of charge carriers, electromagnetic pickup, fluctuating patch potentials, or resistive losses in the surrounding metal and dielectric, among other possible sources (for an extensive discussion of sources of electric field noise, see Ref.~\cite{RevModPhys.87.1419}).

Compared to most ion traps, the large distance ($d \approx 3~\mathrm{mm}$) between the nanoparticle and the nearest electrode makes the particle relatively insensitive to small voltage fluctuations on the surface.
To fully account for the observed heating rate, the electric field noise would need to be of the order $\bar{S}_{\mathrm{EE}} = \bar{S}_{FF}/q^2 \sim 10^{-5}~\mathrm{V}^2 \mathrm{m}^{-2}/\mathrm{Hz}$, where $q \approx 10~e$ is the electric charge on the nanoparticle.
Assuming the field configuration in Appendix~\ref{sec:app_sub_efield}, this corresponds to a voltage noise $\bar{S}_{\mathrm{VV}} \sim 10^{-9}~\mathrm{V}^2/\mathrm{Hz}$ around the particle resonance.
This level of electric field noise is orders-of-magnitude larger than those expected from common sources.
For example, the simplest model of Johnson-Nyquist noise predicts $\bar{S}_{\mathrm{VV}} = 2 k_\mathrm{B} T R(\omega, T) \lesssim 10^{-20}~\mathrm{V}^2/\mathrm{Hz}$~\cite{RevModPhys.87.1419},
where $R(\omega, T) \lesssim 1~\Omega$ is the effective resistance of the whole circuit across the electrodes at frequency $\omega/2 \pi = 50~\mathrm{kHz}$ and temperature $T=295~\mathrm{K}$.

In addition to these estimates, calibration data were taken with highly charged nanoparticles ($\lvert q \rvert \gtrsim 300~e$), where we do not observe a significant excess in the force noise beyond that in Fig.~\ref{fig:noise_budget}.
This allows us to place an empirical upper limit on the electric field noise around the $z$-resonance of $\bar{S}_{\mathrm{EE}} \lesssim 10^{-7}~\mathrm{V}^2\mathrm{m}^{-2} / \mathrm{Hz}$, or on the voltage noise on the nearby electrodes of $\bar{S}_{\mathrm{VV}} \lesssim 10^{-11}~\mathrm{V}^2 / \mathrm{Hz}$.
While the empirical limits above suggest that electric field noise is subdominant to photon recoil in our system,
given the tunability of electric charge on a levitated nanoparticle demonstrated in this work, we note that the impact of electric field noise in future work can be systematically studied through detailed measurements of the charge dependence of the heating rate and noise properties.

Finally, vibration of the experimental apparatus would physically shift the trapping potential and introduce an effective force noise.
Using a commercial accelerometer (Wilcoxon 731A) mounted on the optical table, we measure a seismic noise at the level of $\bar{S}_{aa} \lesssim 10^{-10}~\mathrm{m}^2 ~ \mathrm{s}^{-4} / \mathrm{Hz}$ at $500~\mathrm{Hz}$, with a corresponding force noise $\bar{S}_{FF} = m^2 \bar{S}_{aa} \lesssim 2.3 \times 10^{-45}~\mathrm{N}^2/\mathrm{Hz}$ where $m$ is the mass of the nanoparticle.
The effective force noise, which is about four orders of magnitude below the backaction force noise, serves as an upper limit as the vibrational noise is expected to be highly suppressed in the higher frequencies around the nanoparticle's resonance.

\section{Impulse reconstruction}
\label{sec:app_impulse_recon}
In the impulse calibration procedure discussed in Sec.~\ref{sec:calibration}, the response of a trapped nanosphere is modeled as a high-$Q$ oscillator whose resonant frequency can drift slowly in time.
Based on the position measurement of the oscillator, we would like to estimate the amplitude of impulsive forces acting on it.
An impulse reconstruction algorithm based on optimal (matched) filtering was implemented to extract the amplitude information while accounting for the unknown phase of the oscillator at the time of the impulse~\cite{wang_mechanical_2024, golwala_exclusion_2000}.

For a known signal shape, an optimal filter is a linear estimator of signal amplitude that optimally weights the signal-to-noise power, assuming the noise is Gaussian and stationary.
For each recorded position waveform $z_n$, where $n = 0, 1, ..., N-1$ represents the time index, the optimal filtered amplitude can be written as
\begin{equation}
A_n = \sum^{N-1}_{k=0} \frac{\tilde{s}^*_k \tilde{z}_k e^{2 \pi i k n}}{J_k},
\label{eq:optimal_filter}
\end{equation}
where $\tilde{s}_k$ and $\tilde{z}_k$ are the Fourier transforms of the normalized signal template and $z_n$, respectively.
Gaussian noise with uncorrelated frequency components is described by the power spectral density $J_k$.
Eq.~\eqref{eq:optimal_filter} can be viewed as a convolution of $\tilde{z}$ and the optimal filter $\tilde{\phi} \equiv \tilde{s}^* / J$, evaluated at time offset $t_n$ through the exponential term providing the appropriate phase shift.

For an impulsive force, the signal template $\tilde{s} \propto \chi$ represents the response of an oscillator with mechanical susceptibility $\chi$ to a broadband force.
Around the resonance frequency where the oscillator's motion rises well above the imprecision noise floor, Eq.~\eqref{eq:optimal_filter} can be simplified as
\begin{equation}
A_n \propto \sum^{N-1}_{k=0} \frac{\tilde{z}_k e^{2 \pi i k n}}{\chi_k},
\label{eq:optimal_filter_approximation}
\end{equation}
where we have used $J \approx \bar{S}_{FF} |\chi|^2 \propto |\chi|^2$ with $\bar{S}_{FF}$ being the amplitude of the white force noise.
In this limit, the reconstructed amplitude is equivalent to force that is estimated by dividing out the susceptibility from the measured position waveform.
Eq. \eqref{eq:optimal_filter_approximation} is used to reconstruct the impulse amplitudes for both the calibration and the DM-search data.
To capture the slow drift of mechanical response, the resonant frequency and the corresponding susceptibility is determined in individual $10 \ \mathrm{ms}$-long windows.

\section{Background characterization}
\label{sec:app_background}
\begin{figure}
    \centering
    \includegraphics[width=0.485\textwidth]{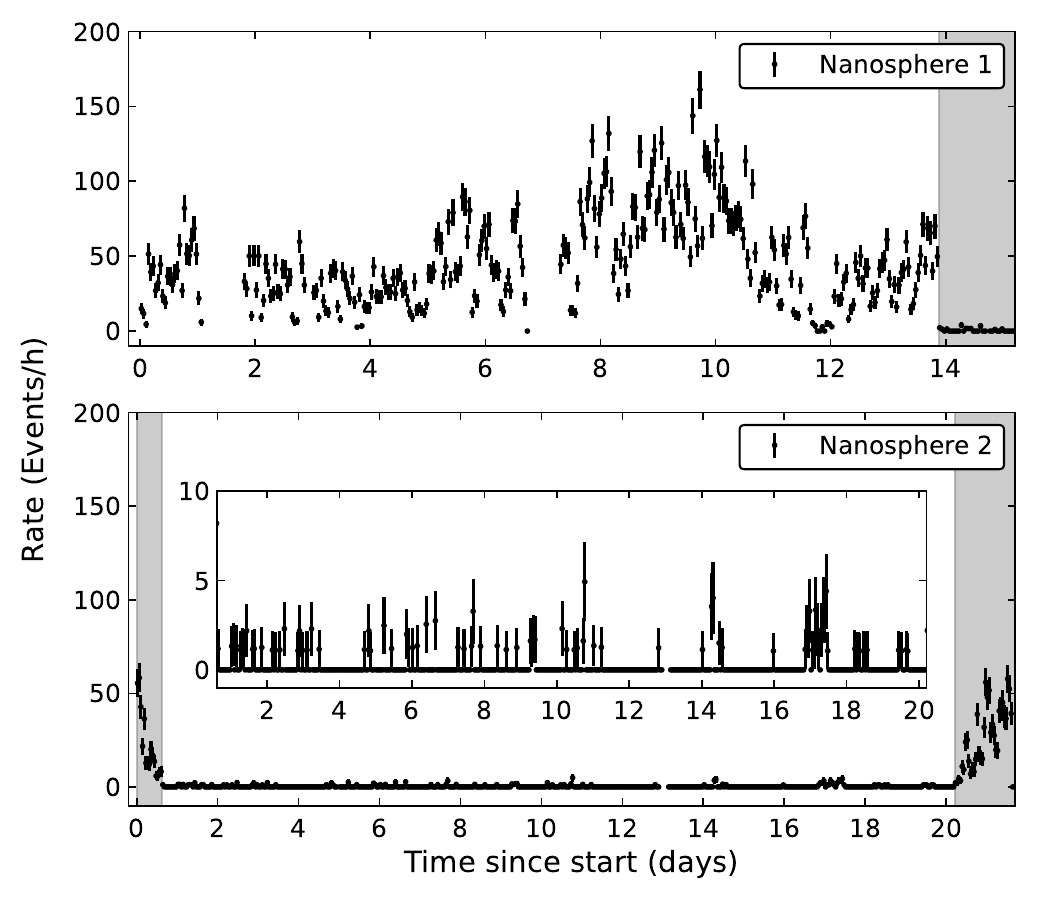}
    \caption{Measured hourly rates for the observed impulse events, including all candidate events that pass the detection quality, noise level, and $\chi^2$ cuts with amplitudes larger than 1.5~MeV/$c$. The error bars represent the Poisson counting errors. Events that happen during the shaded periods show significant deviations from the steady-state rate for each nanosphere, and were excluded from analysis.}
    \label{fig:hourly_rates}
\end{figure}

Impulse events of unidentified origin are observed in the DM-search data described in Sec.~\ref{sec:dm_search}.
Here we study the relative rate of these events as a function of experimental parameters and discuss their possible origin.
Fig.~\ref{fig:hourly_rates} shows the measured hourly rates of large, well-reconstructed ($\geq 1.5~\mathrm{MeV/}c$) events for data measured with Nanosphere 1 and Nanosphere 2.
Significant time correlation in these events is observed in both cases.
For Nanosphere 1, the event rates show a slow modulation on the scale of days, while a considerable reduction near the end of data taking (after $\approx$ 13.9 days since start) was observed, following the removal of a mechanical contact with a mirror mount close to the vacuum chamber.
While not conclusive, this dependence is suggestive that a significant fraction of the previously observed events may have arisen from an uncontrolled coupling of mechanical vibrations to the optical readout.
However, during preliminary investigations of potential background sources, we did not find significant correlations between similar events and signals from the monitoring microphone and accelerometers installed near the experiment.

In the case of Nanosphere 2, data acquired during the beginning (before $\approx$ 0.7 days since start) and the end (after $\approx$ 20.2 days since start) of data taking show substantially higher event rates compared to the rest of the data.
Although there are no obvious sources of environmental coupling during these noisy periods, these data are excluded from DM analysis because DM-induced events are not expected to show strong correlations in time.
Outside of these periods, in certain cases large impulse events were observed to appear in bursts, as shown in the inset in Fig.~\ref{fig:hourly_rates}.

\begin{figure}
    \centering
    \includegraphics[width=0.49\textwidth]{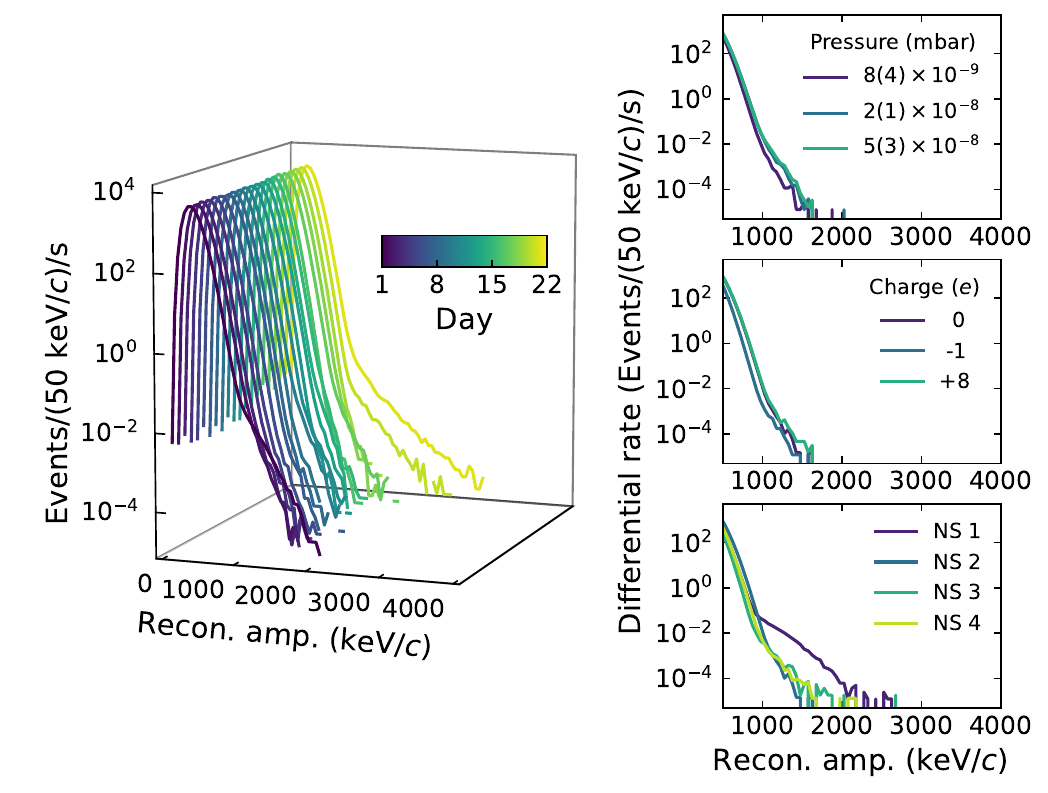}
    \caption{
    Differential event rates measured in various experimental conditions at:
    ({\em Left}) different times throughout data taking, ({\em Top right}) different residual pressures, ({\em Middle right}) different charge states, and ({\em Bottom right}) different nanospheres (NS).
    All spectra, except those labeled as ``NS 1'', ``NS 3``, and ``NS 4'' in the bottom right panel, are taken with NS 2. Each of the spectra represents approximate 24 hours of raw data before data cuts.
    }
    \label{fig:background_characterization}
\end{figure}

Fig.~\ref{fig:background_characterization} shows the full spectra of measured impulse event rates in different experimental conditions for Nanosphere 2 (with the exception of those in the bottom right panel comparing different nanospheres).
Each spectrum represents approximately 24 hours of raw data before selection cuts were applied.
Consistent with Fig.~\ref{fig:hourly_rates}, the observed impulse spectra show strong time-dependent and nanosphere-dependent event rates.
For pressures $\lesssim 10^{-7}~\mathrm{mbar}$, where the force noise becomes independent of pressure, we do not observe significant differences in the event spectra, indicating that the impulses are unlikely to arise from collisions of residual gas molecules.
Furthermore, the amplitudes of these impulses are typically much larger than those expected from gas collisions, even in the high momentum tail of the thermal gas distribution~\cite{barker_collision-resolved_2024}.
However, it remains possible that the observed events could arise from outgassing or desorption of large molecules from the levitated nanospheres~\cite{harvey_nanomechanical_2022}, which is a potential source of decoherence for levitated optomechanics experiments that require a long quantum coherence time~\cite{schafer_desorption-induced_2025}.
We note that while these processes could in principle change the charge state of the nanoparticle, we do not observe such coincident charge changes with the impulse events in the experiment.

Similarly, no significant correlations were observed between the impulse events and the net electric charge states of the nanosphere, suggesting that scattering of stray charged ions, or other transient electric field noise, is unlikely to be responsible for the events.
Besides acoustic and ground vibration, we note that other systematic effects, including nonidealities in the parametric feedback, remain possible sources of these background events that require further investigation.

\section{Calculation of DM event rate}
\label{sec:app_dm_rate}
\subsection{Scattering cross section}
In this work, we focus on DM models that interact with neutrons via a light mediator $\phi$ with mass $m_\phi$.
The effective interaction range is $\lambda \equiv \hbar / m_\phi c$.
Throughout the analysis, we work in the limit where the mass of DM particles, $M_\chi$, is much smaller than the total mass of the levitated nanosphere ($m \approx 4.8~\mathrm{fg} \sim 10^9~\mathrm{GeV}/c^2$).
DM particles are treated as pointlike objects, which is valid as long as the size of DM is much smaller than $\lambda$ and the size of the nanosphere.
In this limit, the problem of DM-nanosphere scattering reduces to an effective one-body problem in the center-of-mass (CM) frame with a reduced mass $M_\chi$ scattering off a fixed spherically symmetric potential.

We generalize Eq. \eqref{eq:yukawa_point_like} and model a levitated nanosphere as a uniform density sphere of radius $R$.
The potential of the nanosphere experienced by a DM particle is given by \cite{monteiro_search_2020, xu_resonant_2021}
\begin{widetext}
\begin{equation}
\label{eq:exact_v}
    V(r) = 
    \begin{cases}
        \frac{3 \alpha}{(R/\lambda)^3} [\lambda^{-1} - \frac{1 + R / \lambda}{r(1 + \coth{(R / \lambda)} )} \cdot \frac{\sinh{(r/\lambda)}}{\sinh{(R/\lambda)}}], &\quad r \leq R, \\

        \frac{3 \alpha}{(R / \lambda)^3} \frac{e^{-r / \lambda}}{r} [\lambda^{-1} R \cosh{(R / \lambda)} - \sinh{(R / \lambda)}], & \quad r > R,
    \end{cases}
\end{equation}
\end{widetext}
where $r$ is the distance to the origin of the CM frame and $\alpha$ is the total coupling strength between the DM particle and the nanosphere.
In the limit $R \rightarrow 0$, the potential reduces to Eq. \eqref{eq:yukawa_point_like}, while in the limit of a massless mediator ($\lambda \rightarrow \infty$), the inverse square law holds exactly and the potential reduces to the case of Rutherford scattering.

In the parameter space studied in this work, the weak coupling limit holds, where small angle scattering dominates the event rates above threshold~\cite{khrapak_momentum_2004, khrapak_classical_2014}.
Without loss of generality, we assume a repulsive potential with $\alpha > 0$ throughout the calculation so a DM particle would never form a bound state with the nanosphere.
The differential cross section $d \sigma /d q$ is numerically solved for classical scattering from this potential~\cite{goldstein_classical_mechanics}.

\subsection{DM velocity distribution}
For DM that is virialized within the Galactic halo, we assume the velocity distribution, $f(v)$, follows a Maxwell-Boltzmann distribution, boosted by the Earth velocity $v_E$, and truncated at the Galactic escape velocity $v_\mathrm{esc}$ \cite{lewin_review_1996}:
\begin{widetext}
\begin{equation}
\label{eq:shm_f_v}
    f(v) = \frac{1}{N_0} \frac{\pi v^2_0}{v_E} \cdot v \cdot
    \begin{cases}
        e^{-(v-v_E)^2 / v^2_0} - e^{-(v+v_E)^2 / v^2_0}, &\quad v + v_E < v_{\mathrm{esc}}, \\

        e^{-(v-v_E)^2 / v^2_0} - e^{-v^2_\mathrm{esc} / v^2_0}, &\quad v + v_E > v_{\mathrm{esc}}, \ v - v_E < v_{\mathrm{esc}}, \\

        0, &\quad \mathrm{otherwise},
    \end{cases}
\end{equation}
\end{widetext}
where $v_0 = 220 \ \mathrm{km/s}$, $v_E = 240 \ \mathrm{km/s}$, $v_\mathrm{esc} = 550 \ \mathrm{km/s}$, and
\begin{equation}
\label{eq:N0_param}
    N_0 = \pi^{3/2} v^3_0 \cdot [\mathrm{erf}(\frac{v_\mathrm{esc}}{v_0}) - \frac{2}{\sqrt{\pi}} \frac{v_\mathrm{esc}}{v_0} e^{-v^2_\mathrm{esc} / v^2_0}]
\end{equation}
is a normalization factor.
In the case that DM is thermalized to a temperature $T$, we assume
\begin{equation}
\label{eq:thermal_f_v}
    f(v) = \frac{1}{N^E_0} \cdot
    \begin{cases}
        4 \pi v^2 e^{-v^2 / (v^T_0)^2}, &\quad v < v^E_{\mathrm{esc}}, \\

        0, &\quad \mathrm{otherwise},
    \end{cases}
\end{equation}
where $v^T_0 = \sqrt{2 k_\mathrm{B} T / M_\chi}$, $v^E_{\mathrm{esc}} = 11.2 \ \mathrm{km/s}$ is the escape velocity of the Earth, and $N^E_0$ is the same as  Eq. \eqref{eq:N0_param} but with $v_0$ and $v_\mathrm{esc}$ replaced by $v^T_0$ and $v^E_{\mathrm{esc}}$.

For the daily modulation analysis in Sec.~\ref{sec:dm_search}, the directional information in the velocity distribution of virialized DM particles is retained, for which~\cite{lewin_review_1996, bozorgnia_daily_2011, mayet_review_2016}
\begin{equation}
    f(\vec{v}) = \frac{1}{N_0} e^{- (\vec{v} +\vec{v}_\mathrm{lab})^2 / v_0^2} \Theta(|\vec{v} +\vec{v}_\mathrm{lab}| - v_\mathrm{esc}),
\end{equation}
where the Galactic DM with velocity $\vec{v}$ is boosted by $\vec{v}_\mathrm{lab}$, i.e., the velocity of the lab with respect to the Galaxy.
The lab velocity is time-dependent and includes components from Galactic rotation, solar motion, Earth revolution, and Earth rotation:
\begin{equation}
    \vec{v}_\mathrm{lab} = \vec{v}_\mathrm{GalRot} + \vec{v}_\mathrm{Solar} + \vec{v}_\mathrm{EarthRev} + \vec{v}_\mathrm{EarthRot}.
\end{equation}
We calculate $\vec{v}_\mathrm{lab}$ at each time of observation following Ref.~\cite{bozorgnia_daily_2011, mayet_review_2016} and determine the projection of the resulting event rate spectra onto the sensor measurement direction, which is measured to be $\mathrm{N} 70 \degree \mathrm{W} \pm 5\degree$ with coordinates 41.3$\degree$N, 72.9$\degree$W, using Monte Carlo simulations.

\section{Atmospheric overburden}
\label{sec:app_overburden}
The velocity distribution of DM particles could be significantly modified on their way to the sensor due to interactions with ordinary matter.
For the measurements presented in this work, which operate at ground level, we expect the atmosphere to be the dominant overburden for the portion of DM incident from the upper hemisphere in the lab frame.
If the DM-neutron interaction is sufficiently strong, the DM particles could be slowed down due to scattering with air molecules and become gravitationally captured by the Earth.

We estimate the momentum transfer involved in DM-molecule scattering following the analogous procedures in the context of plasma physics \cite{khrapak_momentum_2004} and self-interacting DM \cite{tulin_beyond_2013, colquhoun_semiclassical_2021}.
Consider a DM particle at velocity $\vec{v}$ that scatters with an atmospheric molecule at rest via a Yukawa potential $V(r) = \alpha e^{-r / \lambda} / r$, where $\alpha$ and $\lambda$ parametrize the strength and range of the interaction.
Both the DM particle and the molecule are treated as point charges, which is valid as long as the interaction length $\lambda \equiv \hbar / m_\phi c$ is much larger than the physical sizes of both the DM particle and the molecule. For extremely light mediators ($m_\phi \rightarrow 0$), an upper cutoff on $\lambda$ is required to regulate the divergence of the scattering cross section.
However, the cross section diverges only logarithmically in this cutoff, and we find the results are insensitive to the exact details of this cutoff for the parameter ranges considered here.

In the CM frame, the problem reduces to the scattering of a reduced mass $\mu$ with initial velocity $\vec{v}$ from a fixed potential.
With multiple collisions, the direction of $\vec{v}$ does not change on average, while the absolute value decreases in each collision.
The distribution of scattering angle is given by $\sigma^{-1} d \sigma /d \Omega$, where $\sigma$ is the total cross section and $d \sigma /d \Omega$ is the differential scattering cross section per unit solid angle.
The averaged change in the longitudinal velocity per collision is~\cite{khrapak_momentum_2004, colquhoun_semiclassical_2021}
\begin{equation}
    \langle \Delta v_\parallel \rangle = \frac{v}{\sigma} \int \frac{d \sigma}{d \Omega} (1 - \cos \theta) d \Omega = \frac{\sigma_T v}{\sigma},
\end{equation}
where the momentum transfer cross section, $\sigma_T$, is defined as
\begin{equation}
    \sigma_T \equiv \int d \Omega (1 - \cos \theta) \frac{d \sigma}{d \Omega},
\end{equation}
with $\theta$ being the scattering angle in the CM frame. 
The transfer cross section weights the fractional change in momentum by the differential scattering cross section.
In the parameter spaces considered in this work, the classical scattering regime ($\mu v \lambda \gg 1$) with weak coupling ($\beta \equiv \alpha / (\mu v^2 \lambda) \lesssim 1$) is the relevant regime, and the transfer cross section for a repulsive potential is approximately given by \cite{khrapak_momentum_2004, tulin_beyond_2013, colquhoun_semiclassical_2021}
\begin{equation}
    \sigma_T \approx 2 \pi \lambda^2 \beta^2 \mathrm{log}(1 + \beta^{-2}).
\end{equation}
The scattering is azimuthally symmetric so $\langle \Delta v_\perp \rangle = 0$.

Transforming to the lab frame, the average momentum transfer from the DM particle to the air molecule is given by $\langle \Delta p \rangle = \mu \langle \Delta v_\parallel \rangle$.
The rate at which the DM particle loses its momentum while traveling in the atmosphere is then~\cite{colquhoun_semiclassical_2021}
\begin{equation}
    \frac{\dot{p}}{p} = n_\mathrm{atm} v \sigma \langle \Delta p \rangle / (M_\chi v) = \frac{\mu}{M_\chi}n_\mathrm{atm} \sigma_T v,
\end{equation}
where $n_\mathrm{atm}$ is the number density of the atmospheric molecules.

We assume N$_2$ to be the dominant atmospheric gas with an exponentially falling density profile with height.
The average number density is assumed to be $n_\mathrm{atm} = 2.4 \times 10^{25}~\mathrm{m^{-3}}$ in the troposphere with an effective height $h=8.5~\mathrm{km}$.
For DM particles initially traveling at the expected virial velocity, $v \approx 10^{-3}~c$, the DM will take approximately $\Delta t \approx 30~\mathrm{ms}$ to reach the Earth's surface.
For each DM mass, the neutron coupling strength $\alpha_n$ for which $\dot{p} / p = (\Delta t)^{-1}$ indicates the approximate interaction strength where DM particles would be stopped by the atmosphere.
The resulting $\alpha_n$ depends only weakly on the mediator masses for the parameter ranges of interest here, and the case for $m_\phi = 1~\mathrm{eV/}c^2$ is plotted in Fig.~\ref{fig:combined_alpha_limits_virial} and translated to DM-neutron cross sections in Fig.~\ref{fig:composite_dm_limits}.

\bibliography{references}
\clearpage

\end{document}